\documentclass{aa}
\usepackage{graphics} 
\usepackage{epsf} 
\usepackage{epsfig}
\usepackage{psfig}
\usepackage{amssymb}

\usepackage{rotating}
\newcommand{\Msun}{\mbox{\rm M$_{\odot}$}}

\begin{document}

\title{Stellar populations in the CFHTLS\thanks{Based on observations obtained with MegaPrime/MEGACAM, a joint project of
CFHT and CEA/DAPNIA, at the Canada-France-Hawaii Telescope (CFHT)
which is operated by the National Research Council (NRC) of Canada, the
Institut National des Science de l'Univers of the Centre National de la
Recherche Scientifique (CNRS) of France, and the University of Hawaii. This
work is based in part on data products produced at TERAPIX and the Canadian
Astronomy Data Centre as part of the Canada-France-Hawaii Telescope Legacy
Survey, a collaborative project of NRC and CNRS.}}
\subtitle{I. New constraints on the IMF at low mass}

\author{M. Schultheis \inst{1,2}
\and A.C. Robin \inst{1}
\and C.~Reyl\'e \inst{1}
\and H.J. McCracken \inst{2,3}
\and E.~Bertin \inst{2,3}
\and Y. Mellier \inst{2,3}
\and O.~Le F\`evre \inst{4}
}

\authorrunning{Schultheis et al.}

\titlerunning{Stellar populations in the CFHTLS}  

\offprints{mathias@obs-besancon.fr} 

\institute{CNRS UMR6091, Observatoire de Besan\c{c}on, BP1615, F-25010 Besan\c{c}on Cedex France     
\and   CNRS UMR 7095, Institut d'Astrophysique de Paris, 98bis, bld. Arago, 75014 Paris, France  
\and Observatoire de Paris, LERMA 61 avenue de l'Observatoire 75014 Paris, France 
\and Laboratoire d'Astrophysique de Marseille, OAMP, Universit\'e de Provence, UMR 6110, Traverse du Siphon-Les trois Lucs, 1302 Marseille, France 
}

\voffset 1.0truecm

\date{Received ....  / Accepted .. .........}

\sloppy

\abstract{

  We present a stellar populations analysis of the first release of the
  CFHTLS (Canada-France-Hawai Telescope Legacy Survey) data.  A
  detailed comparison between the Besan\c{c}on model of the Galaxy and
  the first data release of the CFHTLS-Deep survey is performed by
  implementing the MEGACAM photometric system in this model using
  stellar atmosphere model libraries. The reliability of the
  theoretical libraries to reproduce the observed colours in the
  MEGACAM system is investigated. The locations of various stellar
  species like subdwarfs, white dwarfs, late-type and brown dwarfs,
  binary systems are identified. The contamination of the stellar
  sample by quasars and compact galaxies is quantified using
  spectroscopic data from the VIMOS-VLT Deep Survey (VVDS) as a
  function of $i'$ magnitude and $r'-i'$ colour.  A comparison between
  simulated counts using the standard IMF at low masses show that the
  number of very low mass dwarfs may have been underestimated in
  previous studies. These observations favour a power law IMF following
  $d(n)/dm \propto m^{-\alpha}$ with $\alpha=2.5$ for $m < $ 0.25
  \Msun\ or $\alpha=3.0$ for $m < $ 0.2 \Msun\ for single stars.  The
  resulting LF is in agreement with the local LF as measured from the 5
  or 25 pc samples. It is in strong disagreement with the Zheng et al
  (2001) LF measured from deep HST data. We show that this discrepancy
  can be understood as an indication of a different IMF at low masses
  at early epochs of the Galaxy compared to the local thin disc IMF.
   }

\maketitle

\section{Introduction}

The CFHTLS (Canada-France Hawaii Telescope Legacy Survey) is a five
year large observing program at the CFH Telescope, using the wide field
prime focus MegaPrime equipped with MEGACAM, a 36 CCD mosaic camera.
Together with its small pixel scale of 0.185\,arcsec and the large
number of nights dedicated to the survey (around 500 nights over five
years), the CFHTLS goes deeper and has a better image quality than the
Sloan Digitized Sky Survey but on a much smaller area of the sky. Hence
it probes a different volume of the Universe.  The scientific goals
from CFHTLS cover a wide range of scales: from the solar system
(systematic search of trans-Neptunian objects), stellar populations and
galactic structure, up to the distant universe, constraining the
geometry of the universe (SNIa and cosmic shear), dark matter
properties, quasars, clusters of galaxies, properties of galaxy
clustering and galaxy evolution at high redshift, and probing the
relation between dark and luminous matter.

\begin{table*}
\caption{Location, median seeing and field of view (in sq. degree) of the 3 CFHTLS fields studied here.} 
\begin{tabular}{ccccccc}
\hline
Field &$\alpha(J2000)$&$\delta(J2000)$&$l (deg)$ &$ b (deg)$ &seeing in $i'$&FOV\\
\hline
D1&02h26m00s&-04d30m00s&172.0&$-58.0$&0.88&0.80 \\
D2&10h00m29s&+02d12m21s&236.8 &42.1&0.95&0.69 \\
D3&14h19m28s&+52d40m41s&96.2&59.6&0.92&0.77 \\
\hline
\end{tabular}
\label{table}
\end{table*}

We plan a series of papers dedicated to analysing stellar populations
in the different surveys components of the CFHTLS.  In this paper we
analyse a subset of the first data release of the CFHTLS, investigating
the objects classified as stellar in the catalogues. We examine the
photometric quality of the three fields of the Deep Survey, D1, D2 and
D3, determine the stellar populations and the contamination by compact
galaxies and quasars.  Using the Besan\c{c}on model of stellar
population synthesis (Robin et al. \cite{Robin2003}) together with
stellar atmosphere models Basel3.1 (Lejeune et al.  \cite{Lejeune97},
Westera et al. \cite{Westera}) and NextGen (Allard et al.
\cite{Allard97}), we are able to produce synthetic star counts,
colour-colour and colour-magnitude diagrams in the MEGACAM filter
system.  We determine the location of white dwarfs, brown dwarfs and
binaries in the different colour-colour diagrams. We emphasize in this
paper the study of the luminosity function at low masses of the disc
population.  A comparison between star counts of the M dwarf
populations in the three CFHTLS Deep fields and model predictions
provide new constraints on the slope of the IMF of low-mass stars, a
parameter which is still under debate.

\section{The CFHTLS fields}

The CFHTLS survey consists of three different surveys:
\begin{itemize}
\item The Deep, a survey comprising four fields (named D1, D2, D3, D4)
  of 1 square degree each, in five filters $u^{*}$,$g'$,$r'$,$i'$ and $z'$
  reaching $r'$ up to $\rm 28^{m}$, observed every 3-4 days,
  
\item the Wide, a survey of 3 patches (W1, W2, W3) each around $\rm
  7\,deg^{2}$ in five bands, with limiting magnitudes $r'$ up to $\rm
  25^{m}$, observed in two epochs separated by 3 years,
  
\item the Very Wide, a survey of $\rm 1300\,deg^{2}$ along the ecliptic
  on 5-6 epochs, dedicated to the detection of fast moving solar system
  objects and stellar populations.
\end{itemize}
 
We refer the reader to the corresponding web pages
(http://www.cfht.hawaii.edu/Science/CFHLS/) for a detailed description
of the survey components.  The observations are carried out at the CFHT
using the MEGACAM camera, which consists of a mosaic of 36 CCDs, each
of them 2048 $\times$ 4612 pixels large. The pixel scale is 0.185''
which gives a total field of view of $0.96 \times 0.94$ deg$^{2}$.  The
seeing of the observations used to construct the stacks is better than
1.1'', which guarantees high quality data for the three different
surveys.

We present in this paper an analysis of three fields of the ``Deep''
Survey, D1, D2 and D3 of the CFHTLS release T0001.  These data are
stacks of many images. Field coordinates and the median seeing in $i'$
in each field are given in Table ~\ref{table}.

\section{Data reduction}

The stacks and catalogues used in this paper were released as part of
the TERAPIX T0001 public release. A brief outline of how these stacks
were prepared is as follows.

CFHTLS observations are carried out with MEGACAM in queue survey mode.
For release T0001, only observations from June, 1st 2003 to July, 22,
2004 were used. Pre-reductions were carried out at the CFHT using the
ELIXIR\footnote{http://www.cfht.hawaii.edu/Instruments/Elixir/}
pre-reduction system at CFHT and then these pre-reduced images were
shipped to TERAPIX via the Canadian Astronomy Data Centre, in Victoria,
Canada. These pre-reduced images were then injected into the TERAPIX
pipeline for inspection and quality control purposes. The TERAPIX tool
QualityFITS was used to inspect and grade each image, and also to
produce weight-maps derived from the CFHT-provided master flats using
the WeightWatcher tool. The global astrometric and photometric
solutions were computed using the WIFIX package, an earlier generation
of the TERAPIX astrometric software, as the production astrometric
software package was still in testing phase at the time of the T0001
release. For inclusion in the stacks, images must have a seeing better
than 1.1'' (1.3'' in $u^{*}$) and airmass less than 1.5.  From this
point on for the image reductions, we followed essentially the same
procedure as outlined in McCracken et al.  (\cite{McCracken2003}), and
we refer to the interested reader to this paper for more details. The
two significant differences are firstly that we use weight maps
computed from the image flat-fields themselves and secondly we use the
USNO-B as the astrometric reference catalogue (which increases the
robustness of the overall astrometric solution with respect to the
solutions utilising the USNO-A).  Full details of the properties of the
final stacks, including depth in each filter and the accuracy of the
final astrometric solution can be found on the TERAPIX web
pages\footnote{http://terapix.iap.fr/article.php?idarticle=383},\footnote{http://terapix.iap.fr/article.php?idarticle=382}. 
The internal accuracy of the astrometric solution (band-to-band) is
better than one pixel rms over the entire MEGACAM field, whereas the
external astrometric solution is around $\sim 0.25"$ rms.

Photometric calibrations for each pre-reduced image is provided by the
ELIXIR pipeline. ELIXIR also applies a CCD-to-CCD flux scaling derived
from repeated observations of dense stellar fields which are shifted
many times around the MEGACAM field of view (providing magnitude
measurements of the same star on different CCDs). This procedure is
necessary to correctly account for the "scattered light" effect and
ensures that the flux of any given object is independent of the
position on the mosaic. The residual ccd-to-ccd magnitude error
following this procedure is around $\sim 3\%$. In constructing the
final stacks, we compare the magnitudes of objects in overlapping
pointings and in each band the photometric exposures are indentified as
those in which the objects have the highest flux: other images are
scaled to these observations.  Based on an examination of galaxy counts
and stellar colour-colour plots (see below), we estimate that our
absolute photometric solution in each filter has a maximum systematic
error of $\sim0.05$~magnitudes. Catalogues were extracted using
SExtractor in dual-image mode, with detections carried out using a
chi-squared image (Szalay et al. \cite{Szalay2003}) constructed from
the $g'r'i'$ images. Kron-like total magnitudes were used throughout.
Through this paper, our magnitudes are presented in the MEGACAM
instrumental AB system.

\section{Star-Galaxy separation}

\begin{figure}
\epsfysize=11.8cm
\centerline{\epsfbox[10 30 590 760]{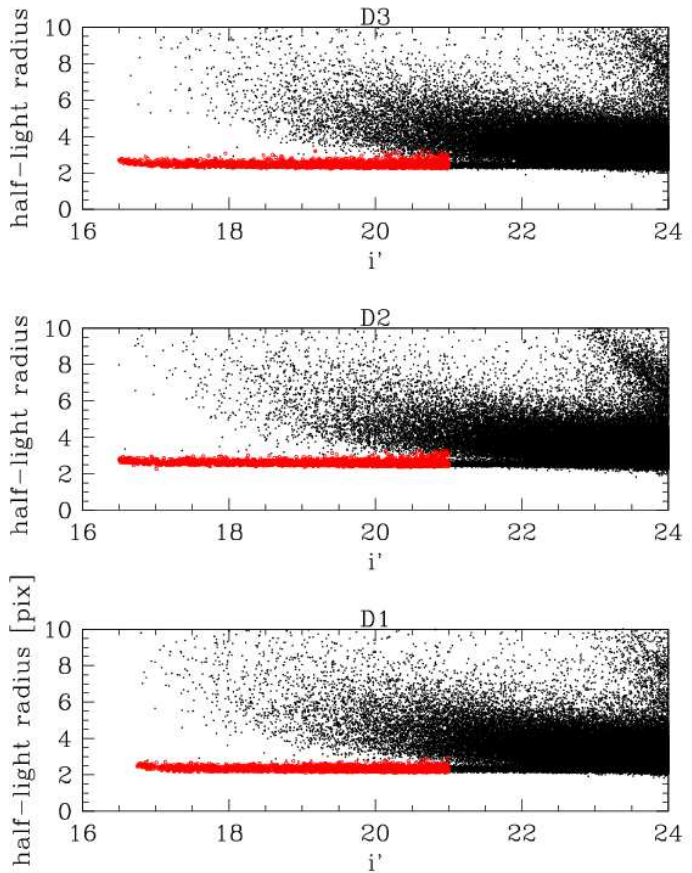}}
\caption{$i'$ magnitude vs half-light radius (in pixels) for the D1, D2 and D3 field. 
The filled circles show the selected stellar branch. Note that objects with $i'<  17$ are saturated.}

\label{star-gal-separation}
\end{figure}

We separated point-like sources from extended ones using SExtractor's
(Bertin \& Arnouts \cite{Bertin96}) "flux-radius" parameter measured on
the $i'$-band image. This parameter measures the radius which encloses
50\% of the object's flux: for point-like sources this is independent
of magnitude, and depends only on the image FWHM.  The stars were
selected by automatically locating the stellar branch in the
flux-radius-magnitude diagram in a series of 10 arc\-minute cells
distributed over each MEGACAM $i'$-stack, which accounts for variation
of FWHM over MEGACAM field of view.  Figure~\ref{star-gal-separation}
shows the compactness parameter against the magnitude for the three
CFHTLS fields D1, D2 and D3. At magnitudes fainter than $i' = 21.0$ the
separation between stars and galaxies starts to be problematic.  From
Fig.~\ref{star-gal-separation} it is clear that for the D1 field the
contamination of galaxies is small for $i' < 21.0$ while for the D2 and
the D3 field the star/galaxy separation starts to fail at already $i'
\sim 20.5$.  The choice of the cut-off at $i' = 21.0$ for the T0001
release is certainly a conservative criterion which can not be applied
for all three fields.  However, the star/galaxy separation depends on
the colour of the objects.  If one restricts to red objects with $r'-i'
> 1.4$, stars can be better seperated from galaxies and thus stars can
be extracted until $i' < 22.5$ (for the D1 field).  We discuss and
quantify below galaxy contamination as a function of magnitude and
colour.

In order to assess the number of galaxies contaminating our stellar
sample, we used spectroscopic data from the VIMOS-VLT Deep Survey
(VVDS).  The VVDS (Le F\`evre et al. \cite{LeFevre2005}) intends to
measure redshifts over $0<z<5$ across 16 deg$^2$ in four separate
fields. The survey is conducted in two steps : an imaging survey and a
spectroscopic survey. Deep imaging in the UBVRI bands and partly $K'$
band is obtained with the CFHT, ESO-NTT and ESO-2.2m. The VLT VIMOS
instruments allow the measurement of redshifts of objects selected from
the imaging survey. The so-called ``Deep'' survey has a limiting
magnitude I$_{AB}$ = 24. Its location, $\alpha$ = 2 h 26 m and $\delta$
= $-4^\circ 30'$ overlaps with that of the D1 field. For a detailed
description of the VVDs data we refer to Le F\`evre et al.
(\cite{LeFevre2005}). We extended the limiting magnitude for the
star/galaxy separation of the D1 field to $i' = 22.0\,mag$ (thus 1\,mag
deeper than the official T0001 release), using the half light radius,
as described above. The D1 stacks have better seeing than the other
fields, allowing star-galaxy separation to fainter magnitudes. We use
this deeper catalogue in order to discuss the galaxy contamination (see
below). However, for the rest of the paper we use the official T0001
release, in which stars are separated from galaxies only until
$i'=21.0$.

\begin{table}
\caption{Galaxy contamination  of the D1 field cross-identified with the VVDS field as a function of $i'$ magnitude
and $r'-i'$ colour. The common area is $\sim$ 0.4 sq degree.}
\begin{center}
\begin{tabular}{|cccc|}
\hline
 $r'-i'$ & $i' < 20.0$&$i'< 21.0$&$i' < 22.0$\\
\hline
 total &3.2\% $\pm 1.8$ & 6.7\% $\pm 1.9$   & 13.1\% $\pm 2.3$\\
 $< 0.5$&  3.2\% $\pm 3.2$   & 13.0\%  $\pm 4.3$   & 23.5\% $\pm 4.9$ \\
$> 0.5$& 3.1\% $\pm 2.2$ & 2.7\% $\pm 1.5$ & 6.5\% $\pm 2.1$ \\
$ > 1.4$& 0\%    &      0\% &     1.8\% $\pm 1.8$ \\
\hline
\end{tabular}
\end{center}
\label{contamination}
\end{table}

\begin{figure}
\epsfxsize=8.5cm
\centerline {\epsfbox[20 20 570 750]{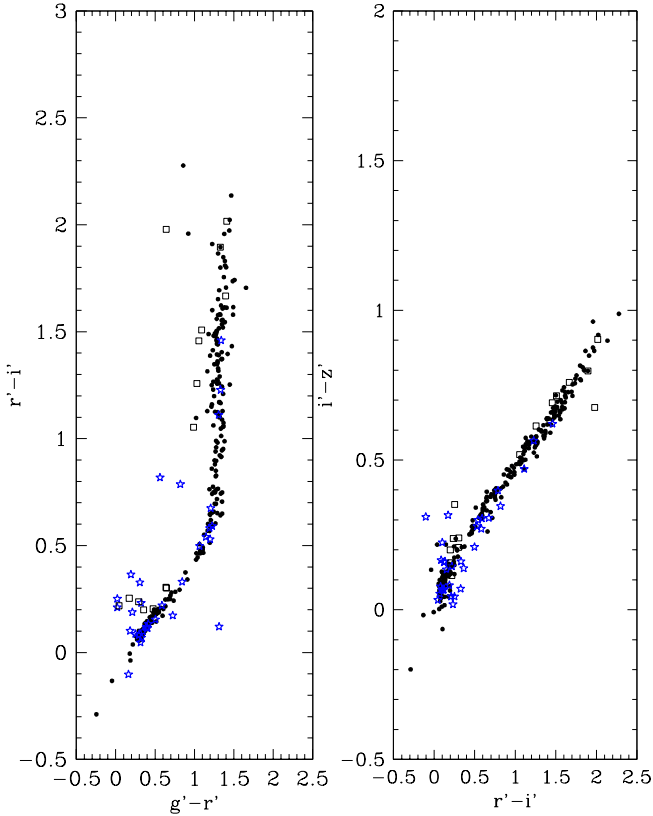}}

\caption{Colour-colour diagrams of the D1 field cross-identified with
  the VVDS data. Objects shown are all morphologically classified as
  stars.  Dots are spectroscopically identified stars. Asterisks are
  galaxies  and open squares indicate
  sources where a blend of a galaxy with a star has been found by
  visual inspection of the images. The limiting magnitude used for the
  morphological star/galaxy separation is $i'= 22.0\,mag$. }
\label{DEEP}
\end{figure}

We cross-identified the D1 data with the VVDS data using a search
radius of 2\arcsec. Due to the masking and only partial overlap between
the D1 field and the VVDS F02-field, only an area of 0.4 sq degree is
in common. Out of 9088 sources of VVDS data, 7110 sources have been
cross-identified with the D1 field. 295 sources were identified as stellar and are shown in Fig.~\ref{DEEP}.

Figure \ref{DEEP} shows the colour-colour diagrams of the D1 field
cross-identified with the VVDS data. All objects shown are classified
as stars from the morphological criterion. The dots show the
spectroscopically identified stars and asterisks those objects
spectroscopically identified as galaxies. The majority of the galaxies
classified morphologically as stars populate the blue part of the
colour-colour diagram with $ r'-i' < 0.4$. They populate the stellar
locus as well as regions outside of the stellar sequence.  Table
\ref{contamination} gives the percentage of galaxies contaminating the
stellar sample as a function of $i'$ magnitude and $r'-i'$ colour.  The
percentage of galaxy contamination depends very much on the $r'-i'$
colour, which means that going to redder colour diminishes
significantly the contamination by galaxies. For $ r'-i' < 0.5$ and $
i' < 22$ one obtains for the D1 field a rate of 23.5\% of contaminating
galaxies while for $ r'-i' > 1.4$ the percentage of galaxy
contamination is negligible. Note that the galaxy contamination is
larger for the D2 and D3 field.

In addition, we visually inspected (for the D1 field) sources which are
located outside the stellar locus.  We have marked them in
Fig.~\ref{DEEP} as open squares.  Most of these objects are galaxies
blended with stars where obviously the aperture photometry is
unreliable. Note that only the $g'-r'$ vs $r'-i'$ diagram reveals these
blends easily.

\section{Stellar atmospheres and the theoretical stellar locus}

\begin{figure*}
  \epsfysize=20.0cm \centerline{\epsfbox[20 17 592 779]{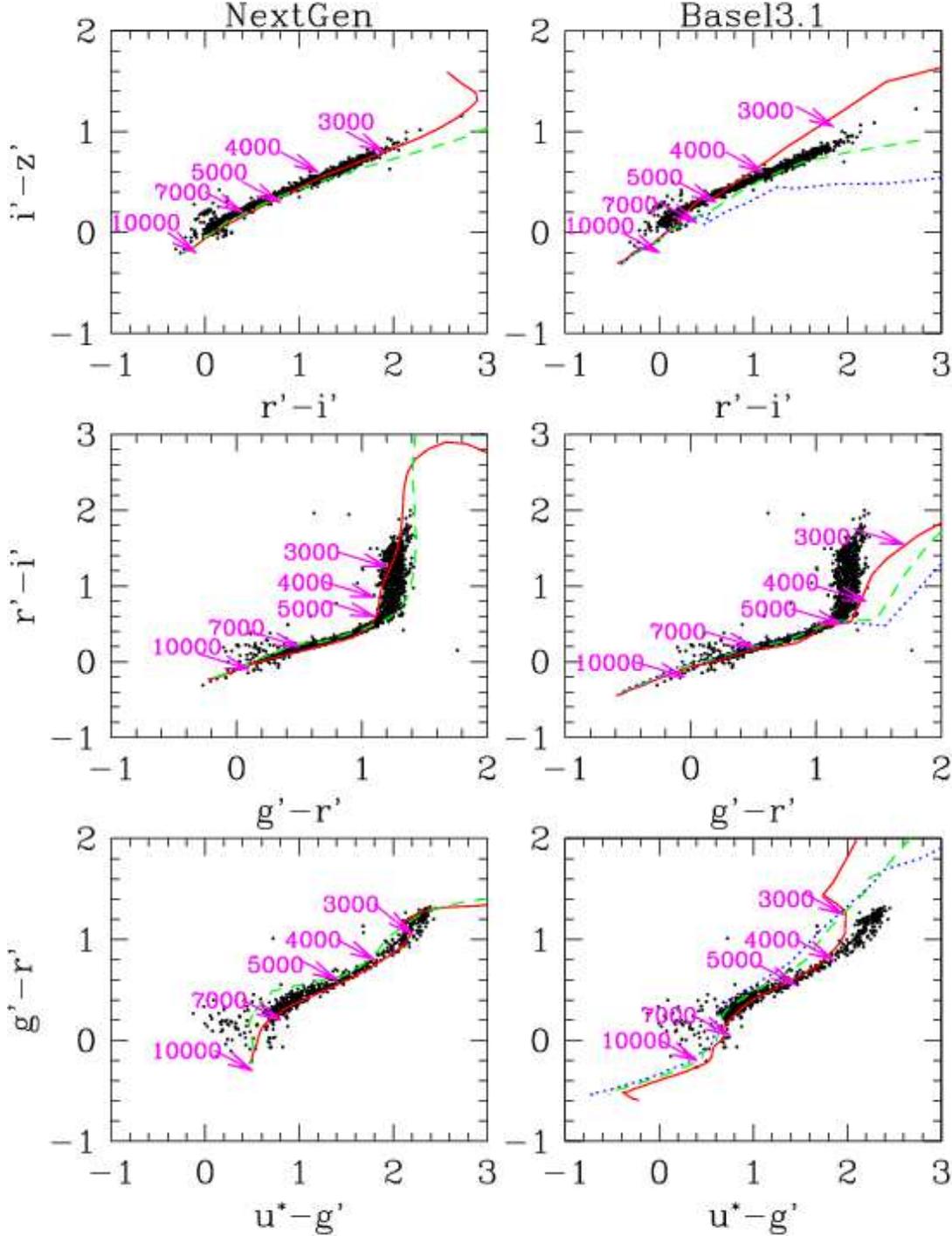}}

\caption{Colour-colour diagrams of the D1 field compared to the
 synthetic colour-colour diagram of NextGen   (left panel) and
 Basel3.1   (right panel) for different metallicities.  The solid line denotes 
[Fe/H]=0.0,  the dashed line  [Fe/H]$=-1.0$ and the dotted line [Fe/H]$=-2.0$. For a better determination of the stellar locus, we show here
  only stellar objects with a photometric error smaller than 0.01\,mag in each filter. The approximate temperatures are also indicated.}
\label{Figure1}
\end{figure*}

Using the published response of the CFHT, Megaprime and MEGACAM we have
computed synthetic colours of the stars as a function of temperature,
gravity and metallicity. To do this we have used two sets of stellar
atmosphere models: Basel 3.1 and NextGen.  The Basel3.1 library is a
semi-empirical library based on preceding generation of models, Basel
2.2 (Lejeune et al.  \cite{Lejeune97}), extended to non-solar
metallicities by Westera et al. (\cite{Westera}). The Kurucz
theoretical spectra (\cite{Kurucz79}) have been modified to fit broad
band photometry using the algorithm by Cuisinier et al. (see Buser\&
Kurucz \cite{Buser92}).  The corrected spectra are used for integrating
the flux in the desired bands. NextGen is the 1997 version of
atmosphere models from Allard et al.  (\cite{Allard97}). These models
use a direct opacity sampling including over 500 million lines of
atomic and molecular species. They give a more realistic description of
the M dwarf population.

Figure \ref{Figure1} shows the CFHTLS colour-colour diagrams of the D1
field superimposed with the synthetic colours of dwarf stars using the
Basel3.1 stellar library for solar metallicity, [Fe/H]$=-1.0$ and
[Fe/H]$=-2.0$ (right panel) and the NextGen library for [Fe/H]=0.0 and
[Fe/H]$=-1.0$ (left panel). For a better definition of the stellar locus,
we use here only stellar objects which have a photometric error
estimate smaller than 0.01\,mag in each filter.  We noted that the $i'-z'$
colour has a slight offset of 0.05\,mag {\bf{ in the D1 field}} compared to the
model which comes from uncertainties on the photometric calibration. For the D2 and D3 field we noted
a shift of 0.07\,mag and 0.02\,mag in $r'-i'$ respectively. 
These offsets have been applied.  

Figure 3 illustrates the sensitivity of the
colours in the CFHTLS system to metallicity and the differences in the
stellar libraries.

For temperatures below 3500\,K, which correspond to K/M stars, the
Basel3.1 library does not give realistic colours for cool dwarfs, but
 gives a better fit to the data than NexGen models do for hotter
stars.

In the temperature range 7000 to 4000\,K, the most sensitive colour is
$g'-r'$, whereas for cooler stars this colour index saturates and $r'-i'$
becomes better. The $i'-z'$ colour seems to be redundant with $r'-i'$, but
going to very cool stars we expect it to be a very good indicator for
selecting brown dwarfs (see sect.7.2).

Both atmosphere models show a strong metallicity effect for cool dwarfs
in $g'-r'$ and $i'-z'$.  It appears that at $g'-r'> 1$ this index is no
longer sensitive to temperature  but  to metallicity. If the
photometric calibration of the survey is accurate and the model
atmospheres reliable, we will be able to constrain the metallicity
distribution of these cool stars, at least statistically. This will permit us to
 determine the metallicity range, and probably the thin disc to thick
disc density ratio, as we expect a difference of metallicity of about
0.5 dex between these two populations, corresponding to about 0.15
magnitude in $g'-r'$ at a temperature of 4000 K.

\section{The Besan\c{c}on Galaxy model}

The Besan\c{c}on Galaxy model is a simulation tool aimed at testing
galaxy evolution scenarii by comparing stellar distributions predicted
by these scenarii with observations, such as photometric
star counts and kinematics. A complete description of the model
ingredients can be found in Robin et al. (\cite{Robin2003}). We
summarise here the model's principal features.

The model assumes that stars are created from gas following a star
formation history and an initial mass function; stellar evolution
follows evolutionary tracks. To reproduce the overall galaxy formation
and evolution we distinguish four populations of different ages and
star formation history, which we now describe.

\begin{figure}
\epsfysize=12cm
 \centerline {\epsfbox[40 20 570 750]{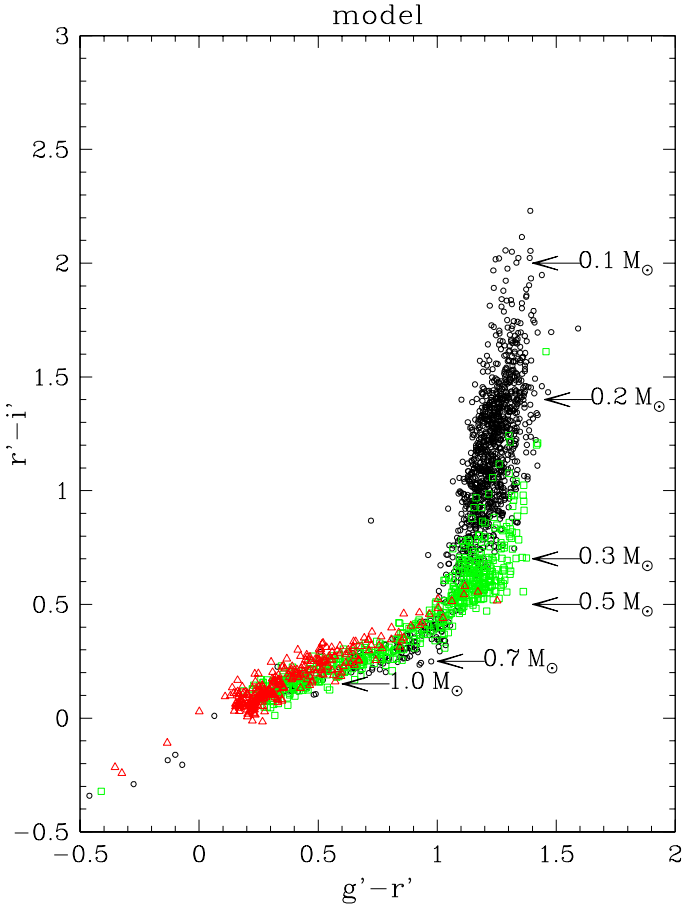}}
\caption{Colour-colour diagram  predicted by the Besan\c{c}on model  for the D1 field  for thin disc (dots), thick disc  (open squares) and spheroid  (asterisks) stars.} 
\label{ccd-components}
\end{figure}

The oldest population is the spheroid. For this population we assume a
single-burst star formation history ocurring early in the
lifetime of the Galaxy, around 14 Gyr ago, from gas still in a
spheroidal distribution. The initial mass function (IMF) and the
density distribution of this population is characterised by a power law
index, an axis ratio and a local normalisation, which are constrained
by remote star counts at high and medium Galactic latitudes (Robin et
al.  \cite{Robin2000}).  The kinematics are also deduced from in situ
velocity determinations.  The mean metallicity of the spheroid is
assumed to be --1.7\,dex with an intrinsic dispersion of 0.5\,dex.

Secondly, a population, called the thick disc, is formed of stars born
about 11-12 Gyr ago in a short period of time as implied by recent
metallicity determinations for this population. We also assume a single
burst for simplicity.  For the thick disk, star formation occurred from
the gas already settled in the disk.  The kinematics, deduced from
observational constraints (Ojha et al. \cite{Ojha96}, Ojha
\cite{Ojha99}), implies that it has undergone a merging event shortly
after the disc formation (Robin et al. \cite{Robin96}), increasing the
disk thickness and giving a higher velocity dispersion and scale
height.  The IMF, density distribution and local normalisation were
constrained from star counts (Reyl\'e \& Robin \cite{Reyle2001}). The
mean metallicity of the thick disc is assumed to be --0.7\,dex with an
intrinsic dispersion of 0.3\,dex.

Thirdly, a bulge population is present in the center of the Galaxy and
extends to about 2 kpc. Its age is of the order of 10 Gyr. This value
is however poorly constrained. This population has a triaxial
distribution, as a bar as determined by Picaud et al.
(\cite{Picaud2004}) from near-infrared star counts. Velocity
dispersions are large, similar to the spheroid. The mean
metallicity is assumed solar with a dispersion of 0.5\,dex.

A standard evolution model is used to compute the disc population,
based on a typical set of parameters: an IMF, a star formation rate
(SFR), a set of evolutionary tracks (see Haywood et al., 1997a and
references therein).  The disc population is assumed to evolve during
10 Gyr.  A set of IMF slopes and SFR's are tentatively assumed and
tested against star counts. The tuning of disc parameters against
relevant observational data has been described in Haywood et al.
(\cite{Haywood97}, \cite{Haywood97a}).

A revised IMF has been used as a starting point in the present
analysis, adjusted to agree with the most recent Hipparcos results: the
age-velocity dispersion relation is from G\'omez et al.
(\cite{Gomez97}), the local luminosity function is from Jahrei\ss\ \&
Wielen (\cite{Jahreiss97}) and an IMF is adjusted to it, giving an IMF
slope $\alpha$ = 1.5 in the low mass range [0.5-0.08 \Msun], in good
agreement with Kroupa (\cite{Kroupa2001}).  The scale height has been
computed self-consistently using the potential via the Boltzmann
equation. The local dynamical mass was taken from Cr\'ez\'e et al.
(\cite{Creze1998}).

The evolutionary model fixes the distribution of stars within the
parameter space of effective temperature, gravity, absolute magnitude,
mass and age. These parameters are converted into colours in various
systems through stellar atmosphere models.

In the case of the MEGACAM photometric system, we have used the
optics, CCD and filter definition of the passbands, and applied them to the
spectral libraries.  As seen in Fig.~1, the Basel3.1 library is more
suitable for hot stars, giving better predicted colours, especially
$u^{*}-g'$, while the NextGen library is more realistic for cool stars.
Hence we have adopted a combination of both: Basel3.1 at $\rm T_{eff} >
4000\,K$ and NextGen for cooler stars. As the cooler stars are mostly
located in the disc, we used only [Fe/H]=0.0 and [Fe/H]$=-1.0$ for $\rm
T_{eff} < 4000\,K$.  The model simulations also include a model of
extinction and account for observational errors. The Besan\c{c}on model
predictions in the MEGACAM photometric system can be found at
http://www.obs-besancon.fr/modele.

\section{Stellar populations}

Simulations from the Besan\c{c}on model in the MEGACAM photometric
system have been performed and compared with the CFHTLS in three
fields: D1, D2 and D3.

We used the same magnitude limit in $i'$ of 21.0 imposed by the
star/galaxy separation and took the photometric errors in each band as
a function of magnitude into account. In the discussion below we
indicate also the different components of the Galactic model
such as thin disc, thick disc and the spheroid.  Figure
\ref{ccd-components} shows the expected colour-colour diagram predicted
by the Besan\c{c}on model for the D1 field. Indicated are the three
different components, thin disc, thick disc and halo. While the blue
part of the colour-colour diagram is populated by mainly spheroid stars,
the thick disc stars are concentrated around $0.6 <r'-i'< 0.9$ and
the thin disc stars cover the red part of the diagram.

Figure \ref{ccd-all} shows the colour-colour diagrams of the CFHTLS for
the D1, D2 and D3 fields respectively as well as the model predictions.
Note the excellent overall agreement between observed and predicted
colours for the three Galactic components.  Figure ~\ref{ccd-all} also
shows that the stellar populations of the three deep fields are
similar, although one notices several differences:

\begin{figure*}
\epsfysize=20.5cm
 \centerline {\epsfbox[10 20 570 750]{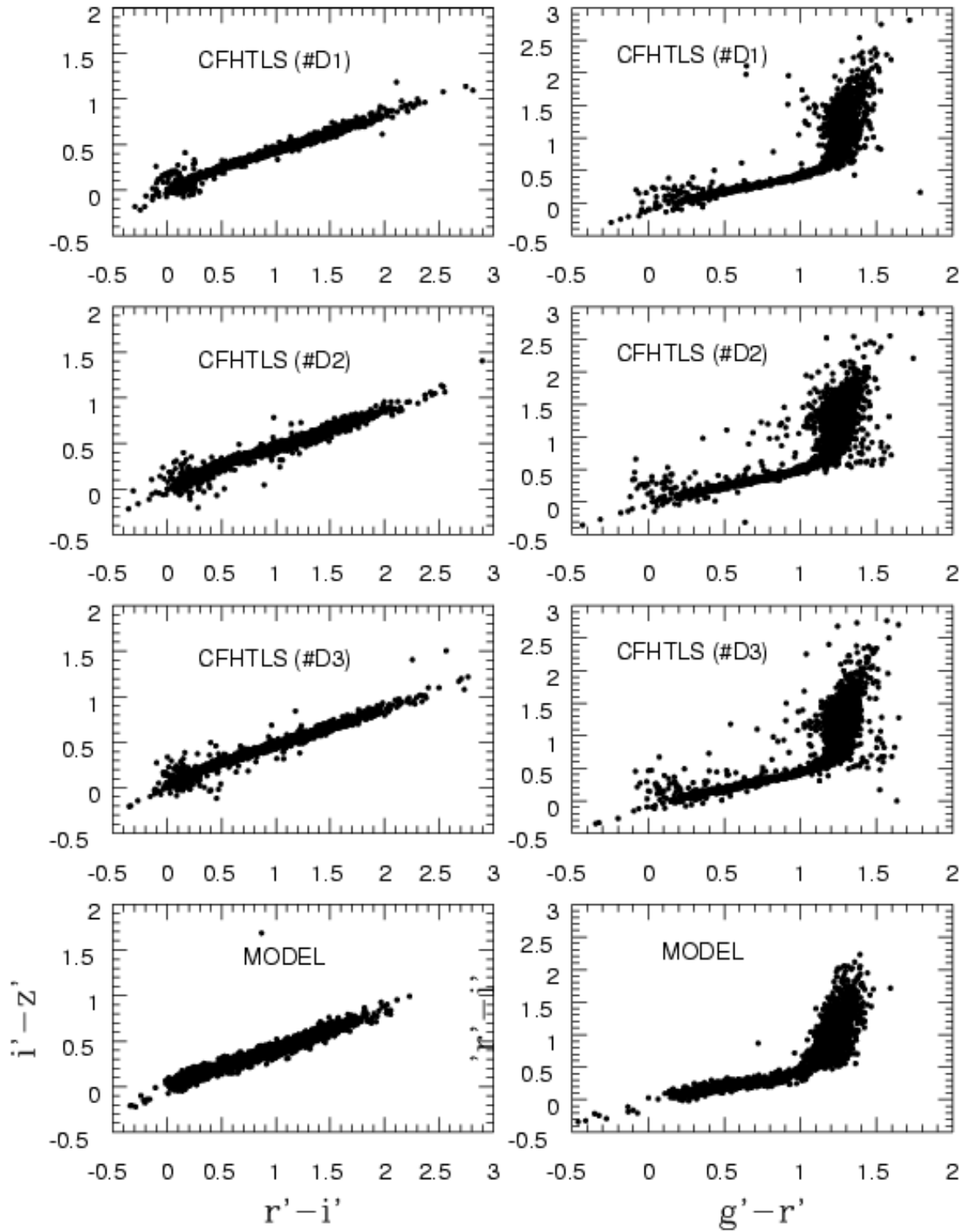}}
\caption{Colour-colour diagrams of the three Deep field D1, D2 and D3  compared with
        synthetic colours predicted by the Besan\c{c}on model (lower panel). Note the excellent agreement between the data and the model. }
\label{ccd-all}
\end{figure*}

\begin{itemize}
  
\item there is a steady increase in the width of the stellar locus from
  the D1 field - which has a well defined narrow locus - towards the D3
  field - which is wider - and the D2 field which is wider still,
  especially at $r'-i' > 1.2$, and more pronouced in $i'-z'$.

\item The blue part of the colour-colour diagrams ($g'-r' < 0.5$, $r'-i'
  < 0.4$, $i'-z' < 0.3$) is populated by an overdensity of objects
  which has a larger spread than the predicted stellar locus.  These
  objects are compact faint galaxies or quasars erroneously classified
  as stars by morphological criteria (see Sect.~4). This contamination
  is larger in D2 and D3 than in D1.
  
\item In the $g'-r'$ vs $r'-i'$ diagram, there are sources with
  $g'-r' > 1.3$ and $r'-i' < 1.5$ abundant in the D2 and D3
  fields, but absent in the D1 field. Pello (priv. communication)
  provided us with theoretical colours of galaxies in the MEGACAM
  photometric system for different redshifts and for different
  morphological types which show that these objects are expected to be
  elliptical galaxies with $z < 0.5$ while a few high redshift QSOs
  and spiral galaxies could also be present.

\item In the D1, D2 and D3 fields there are objects with $r'-i' > 2$
  and $i'-z' > 1.2$ which are off the stellar sequence which may be
  brown dwarf candidates (see Sect.~7.2), or high redshift quasars.
  
\item The dispersion in the synthetic colour-colour diagram is larger
  than observed, especially in the blue part.  This is due to
  uncertainties of the model atmospheres in this metallicity range, as
  seen in Fig.~3.
\end{itemize}

\subsection{White dwarfs}

White dwarfs (WD) are the last stage of stellar evolution and their space 
density depends on the Galactic star formation history and initial mass
function. While the luminosity function (thereafter LF) of white dwarfs
in the thin disc is known from systematic searches in the solar
neighbourhood (e.~g. Liebert et al. \cite{Liebert88}, Ruiz \& Bergeron
\cite{Ruiz2001}, Holberg et al. \cite{Holberg2002}), only very few
thick disc white dwarfs have been identified while the presence of
white dwarfs in the Galactic halo is still uncertain. Knowledge of
the luminosity function of thick disc and halo white dwarfs is expected
to constrain the age of these populations, the physics of the coolest
white dwarfs, as well as the initial mass function at early epochs in
the Galaxy (Chabrier \cite{Chabrier2003}).

Cr\'ez\'e et al. (\cite{Creze2004}) used two-epoch observations of
the $\rm 1\,deg^{2}$ VVDS-F02 deep field to search for white dwarfs in the VVDS
survey by proper motions; they reported a null detection. 

\begin{figure}
\epsfxsize=8.5cm
\centerline{\epsfbox[100 80 470 780]{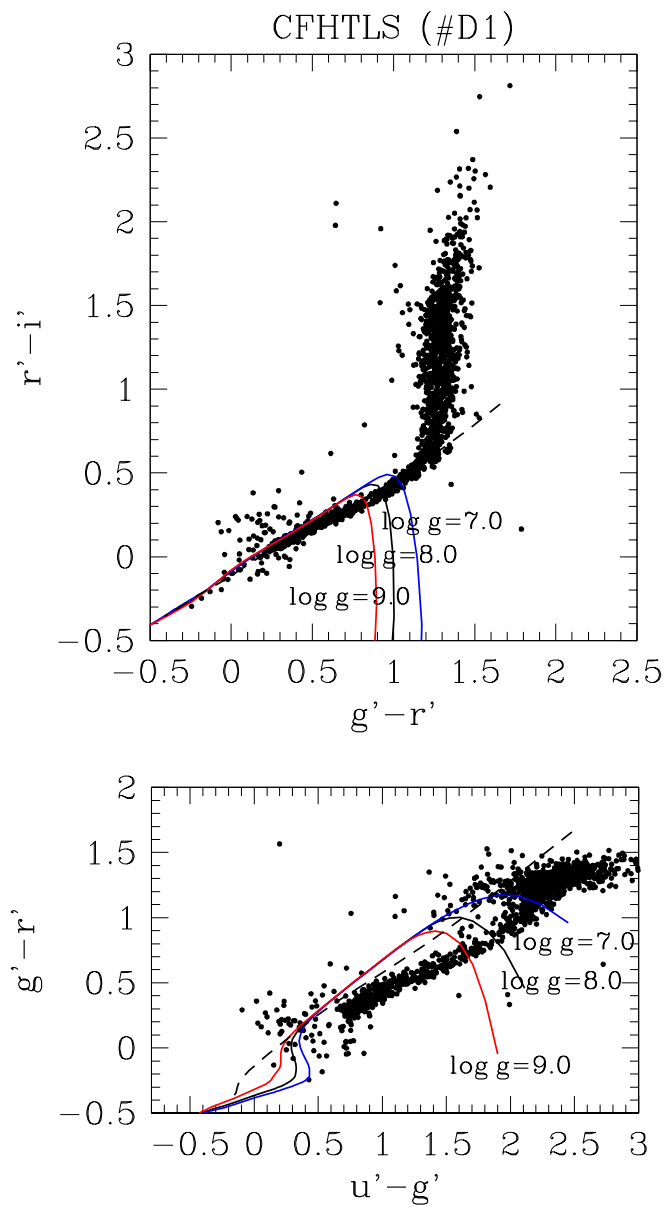}}
\caption{White dwarf sequences in colour-colour diagrams of the D1 field. Solid line shows the DA sequence, dashed line the DB sequence from theoretical model from Bergeron et al. (\protect\cite{Bergeron01}) for surface gravities of log\,g=7.0, 8.0  and 9.0}
\label{wd}
\end{figure}

Bergeron (priv. communication) provided us with theoretical colours of
white dwarfs in the MEGACAM photometric system.  Figure~\ref{wd} shows
their location in various colour-colour diagrams compared with data in
the D1 field. The locus of the white dwarfs is distinct from the locus
of the subdwarfs in the $u^{*}-g'$ vs $g'-r'$ plane only. Furthermore,
we note the sensitivity of the colours to log\,$g$. However this part
of colour-colour space is highly contaminated by quasars. In the other
colour combinations, white dwarfs are indistinguishable from subdwarfs,
except the very cool ones (temperature less than about 3200 K) where DA
white dwarfs with hydrogen atmospheres start to deviate from blackbody.

In the present data, there are a few objects with colours consistent
with those of white dwarfs.  However these objects could also be
horizontal branch spheroid stars or quasars. White dwarf candidates
will only be reliably identified when proper motions become available.
Proper motions will easily distinguish these objects from horizontal
branch stars, as they are brighter, hence much more distant at a given
apparent magnitude (extragalactic objects such as quasars, of course,
will have no measurable proper motions). The CFHTLS will allow us to
eventually cover about 150 square degrees to the same magnitude limit
with proper motions, allowing a definitive answer if baryonic matter is
present in galactic halos in the form of white dwarfs.

The number of expected white dwarfs per square degree, as predicted by
the Besan\c{c}on Galaxy model is about 25 for the thin disc, around two
for the thick disc and about 0.1 for the halo, to magnitude $i'=22.5$.
The number of ancient halo WDs is computed assuming that the dark halo
is partly made of ancient white dwarfs, at the level of 2\% of its mass
density. Even at these bright magnitudes their identification from
photometry only will be difficult due to the large number of galaxies
and quasars; proper motions are necessary.  Assuming that the survey
would reach an astrometric accuracy of about 0.1 pixel at $i'=22.5$,
with a time baseline of three years, objects having a motion of 25
mas/yr would be detectable at the 3-$\sigma$ level.  About 2/3 of the
disc WD pass this proper motion selection criterion, 80\% of the thick
disc WDs, and 100\% of the halo WDs.

\subsection{M, L, and T dwarfs}

Large sky surveys are very efficient in identifying extremely late-type
stellar and sub-stellar objects because they are detectable only at
short distances from the sun, and hence in a small volume. Even if they
are numerous, the number of brown dwarfs detectable to date is still
small.  Near-infrared surveys such as DENIS and 2MASS resulted in the
detection of a significant number of close low-mass stars and brown
dwarfs (see e.~g. Kendall et al. \cite{Kendall2004}, Reid et al.
\cite{Reid2004}, Burgasser et al. \cite{Burgasser2004}). The SDSS
survey was used to identify the first field T dwarfs (Strauss et al.
\cite{Strauss99}, Tsvetanov et al. \cite{Tsvetanov2000}) and the
numbers of known L dwarfs has been greatly increased(Fan et al.
\cite{Fan2000}, Schneider et al. \cite{Schneider2002}).  Gelino et al.
(\cite{Gelino2004}) introduced an homogeneous database of M, L, and T
dwarfs, that contains more than one thousand of these objects.  Hawley
et al.  (\cite{Hawley2002}) compiled a large sample of M, L and T
dwarfs from SDSS spectra together with SDSS photometry and additional
near-IR photometry (2MASS). They find that the $i-z$ and $i-J$ colour
are the most useful for estimating spectral types based solely on
photometric information for M and L dwarfs.

\begin{table*}
\caption{Positions and MEGACAM photometry of late type dwarfs or high redshift quasars.}
\begin{tabular}{|cccccc|}
\hline
Field& ra (J2000) & dec (J2000) & $i'$(mag) & $r'-i'$& $i'-z'$\\
\hline
D1&2h24m58.2s &$-4^{0}$ 17\arcmin 00.3\arcsec& 20.43&2.11&1.28\\
D2&9h59m4.1s &+2$^{0}$34\arcmin 27.9\arcsec & 19.66 &2.90&1.40\\
D3&14h19m53.4s&+52$^{0}$17\arcmin52.9\arcsec&20.39 &2.56&1.51 \\
D3&14h22m01.6s&+52$^{0}$48\arcmin46.6\arcsec&20.57 &2.26&1.40\\
\hline
\end{tabular}
\label{naines_candidates}
\end{table*}

\begin{figure}
\epsfxsize=8.5cm
\centerline{\epsfbox[60 160 540 610]{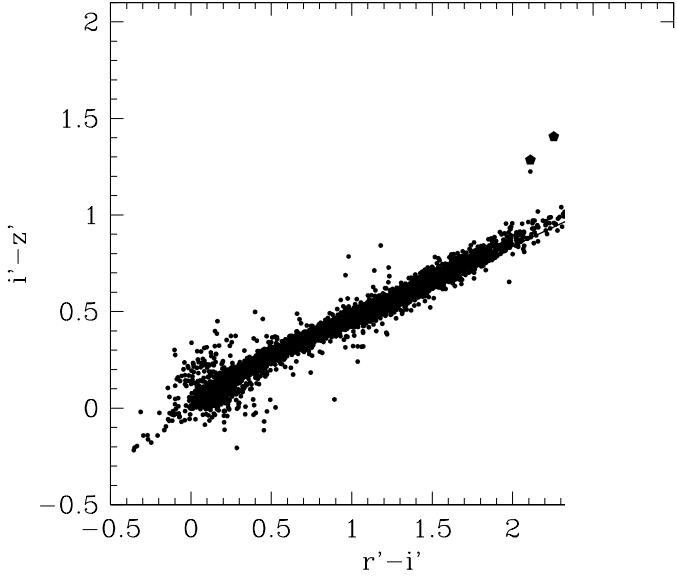}}
\caption{$r'-i'$ vs $i'-z'$ diagrams of the D1, D2 and D3 fields. Superimposed are the average colors of M and L dwarfs taken from NextGen models. Spectral types are
indicated, following the temperature-spectral type relation from Golimowski et al. (\protect \cite{Golimowski2004}). Candidate L dwarfs are shown as
asterisks (see also Tab.~3).}
\label{naines}
\end{figure}

Figure \ref{naines} shows the $i'-z'$ vs $r'-i'$ diagram of the three
CFHTLS fields together superimposed with the NextGen model at solar
metallicity.  The spectral types are from the temperature-spectral type
relation from Golimowski et al. (\cite{Golimowski2004}). There are two
objects with $r'-i' > 2$ and $i'-z' > 1.3$ in the D3 field and one
candidate each in the D1 and D2 fields, which are good candidates for
being either early L dwarfs or high redshift quasars (see
Fig.~\ref{naines}). They are given in Tab.~3. High redshift quasars
with $i'-z' >1.5$ will be distinguishable from brown dwarfs either by
near-infrared photometry or by proper motion measurements.

\subsection{Binary systems}

The formation and evolution of low-mass stars in a binary system is a
common phenomenon which leads to the interesting class of cataclysmic
variables. In deep surveys one expects to detect a few cases of WD-M
dwarf pairs. Raymond et al. (\cite{Raymond2003}) identified $\sim$ 100
white dwarf-M dwarf pairs in the SDSS survey with $g < 20$. Using
additional spectroscopy, they achieve an efficiency of $\sim 60\%$
in finding white dwarf-M dwarf pairs because of the contamination by
galaxies in the interesting colour regions.

\begin{figure}
\epsfxsize=8.5cm
\centerline{\epsfbox[108 160 460 730]{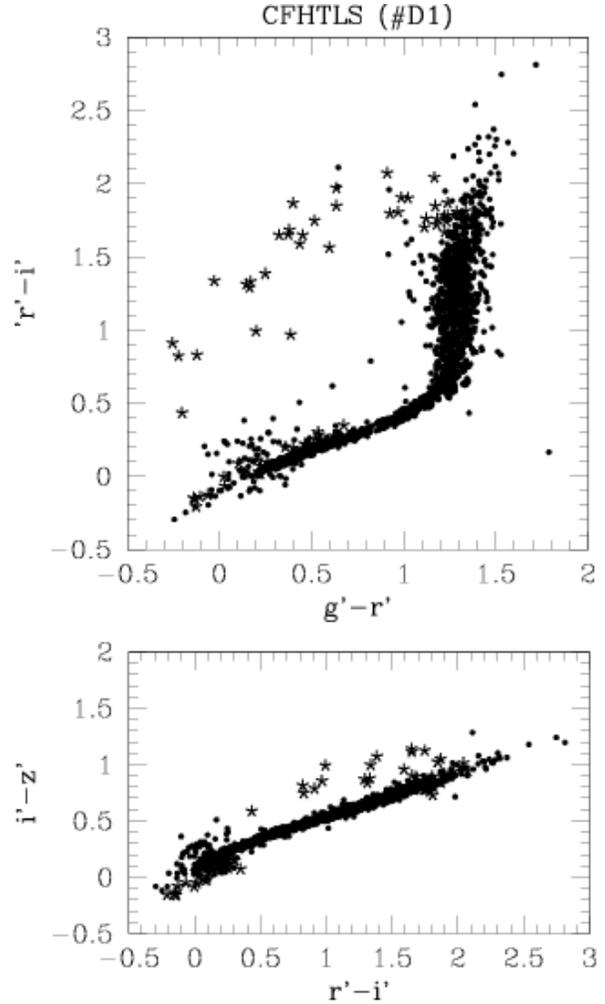}}
\caption{Binary systems M dwarf + White dwarf in the D3 field. Dots: whole D1 data set. Stars: simulated systems. Colours are 
estimated by adding up the flux in a realistic sample of unresolved systems.}
\label{binaires}
\end{figure}

We simulated a sample of unresolved M dwarfs + white dwarfs systems by merging their fluxes.
 Typical colours of these simulated systems are given in 
Fig.~\ref{binaires} as star symbols. The location of these systems is 
clearly outside the single star locus in the $g'-r'$ vs $r'-i'$ diagram. However 
they lie in a region where we expect contamination by compact galaxies and 
quasars. Their identification will be easy using proper motions, all these 
objects being intrinsically faint, and are hence detected only in the solar 
neighbourhood.

\section{Stellar densities and the  IMF at low masses}

\begin{figure}[h!]
\epsfxsize=8.5cm 
\centerline{\epsfbox[20 20 570 750]{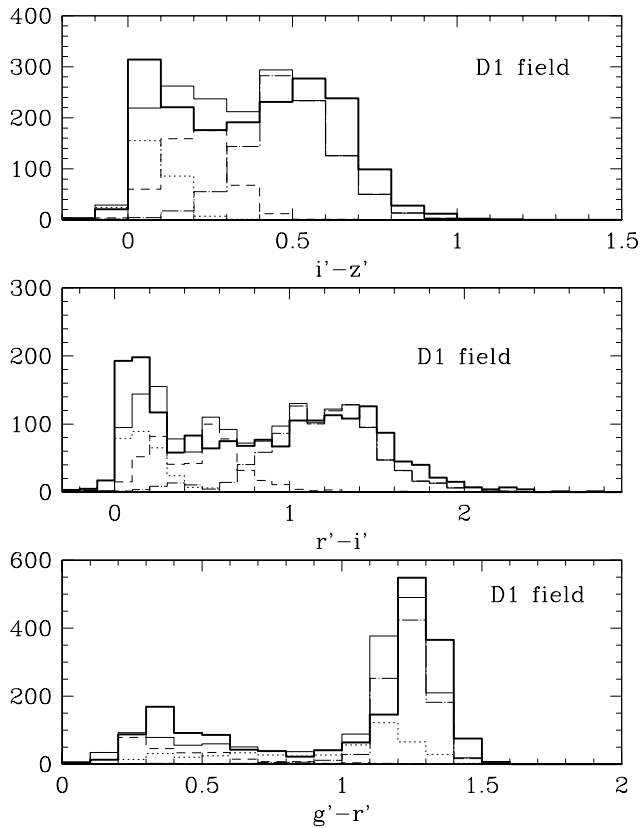}
}
 \caption{Histograms of the colour distributions in the D1 field. The thick  line denotes the observations,
 the thin line the Besan\c{c}on model, the dotted line the contribution of the spheroid, the short-dashed line the thick disc and the dot-long dashed line the thin disc contribution.}
\label{histD1}
\end{figure}

\begin{figure}
\epsfxsize=4.4cm
\centerline{\epsfbox[35 370 530 730]{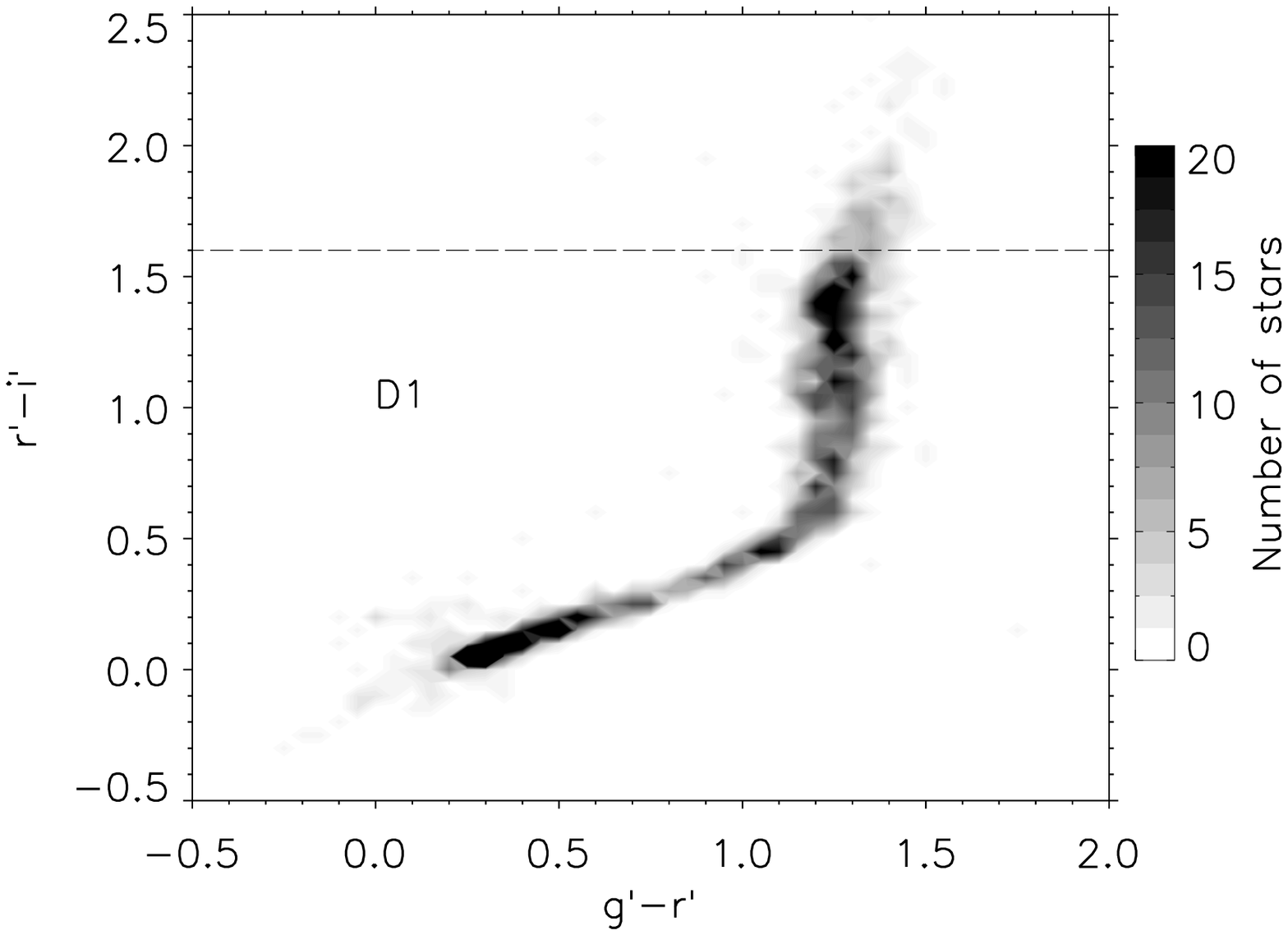} \epsfxsize=4.4cm \epsfbox[35 370 530 730]{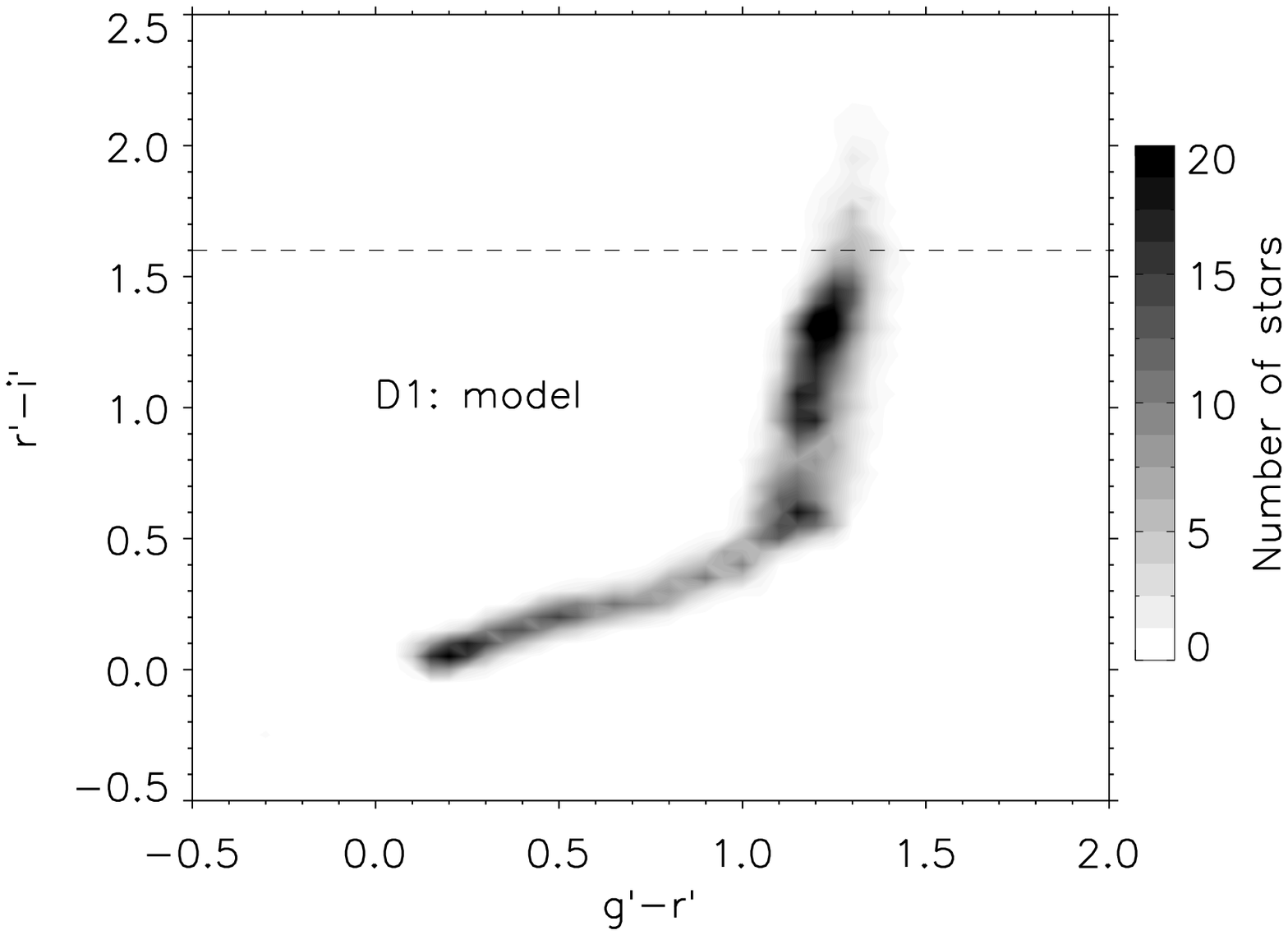}} 
\centerline{\epsfbox[35 370 530 730]{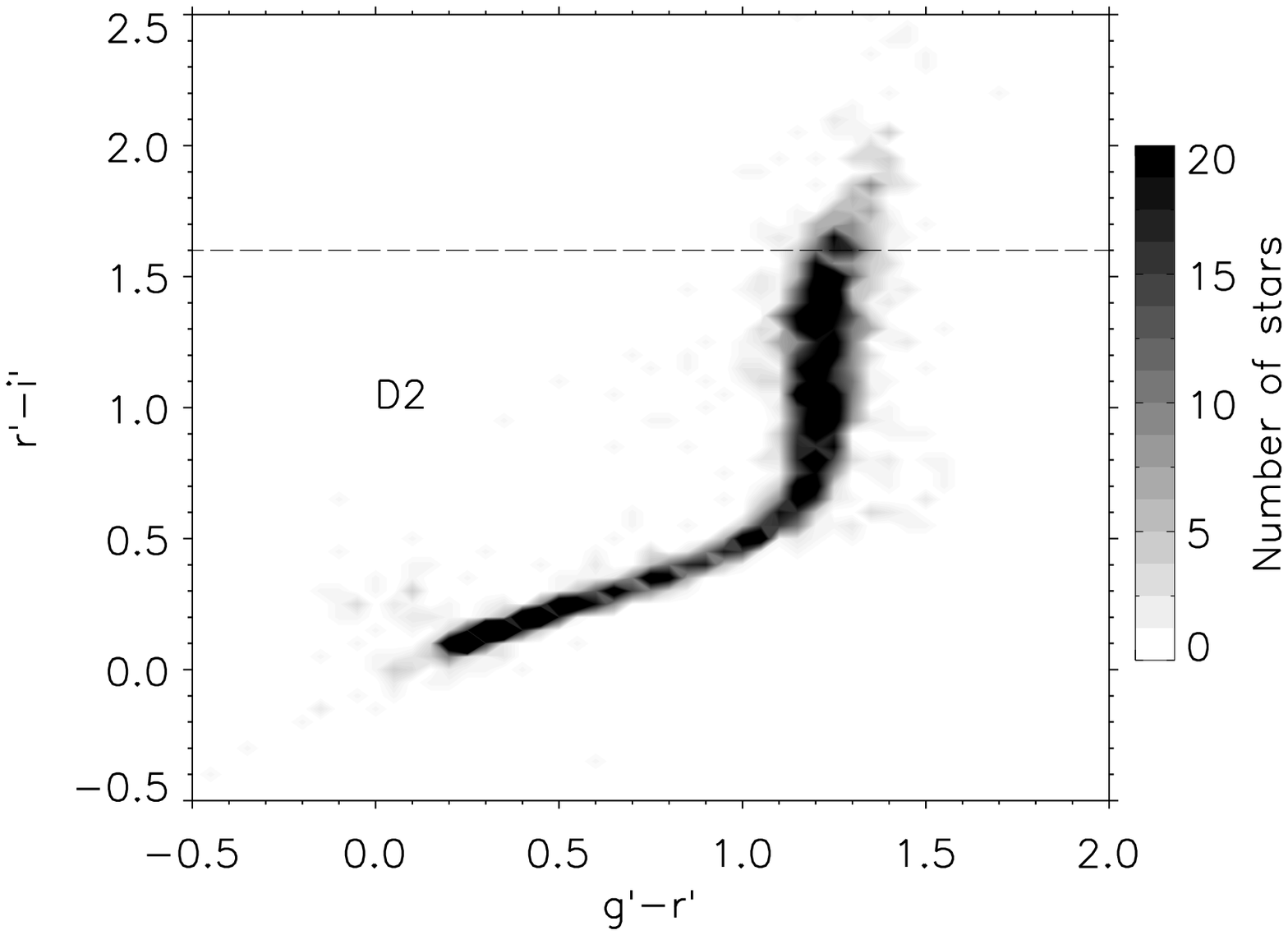} \epsfxsize=4.4cm \epsfbox[35 370 530 730]{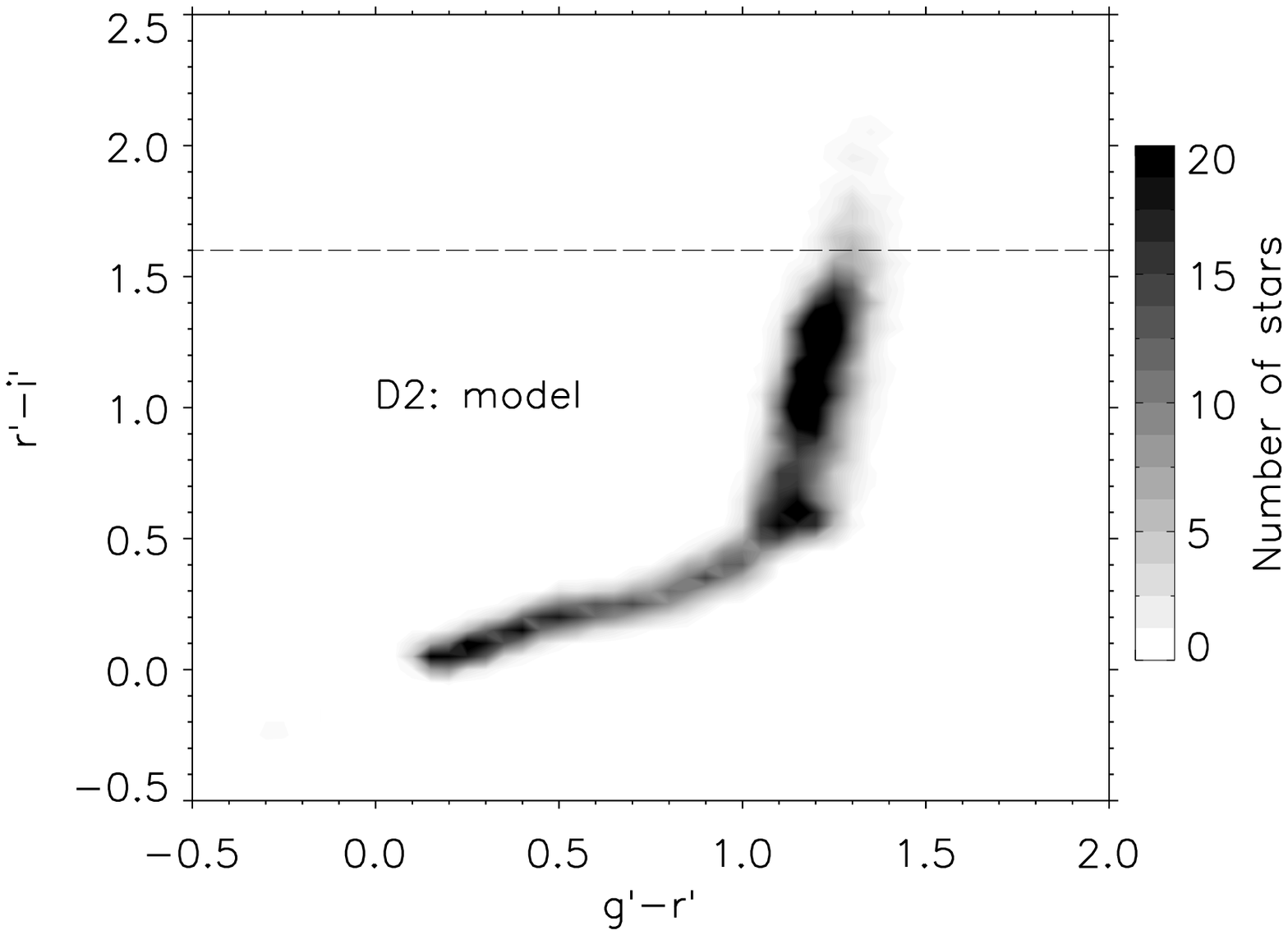}}
\centerline{\epsfbox[35 370 530 730]{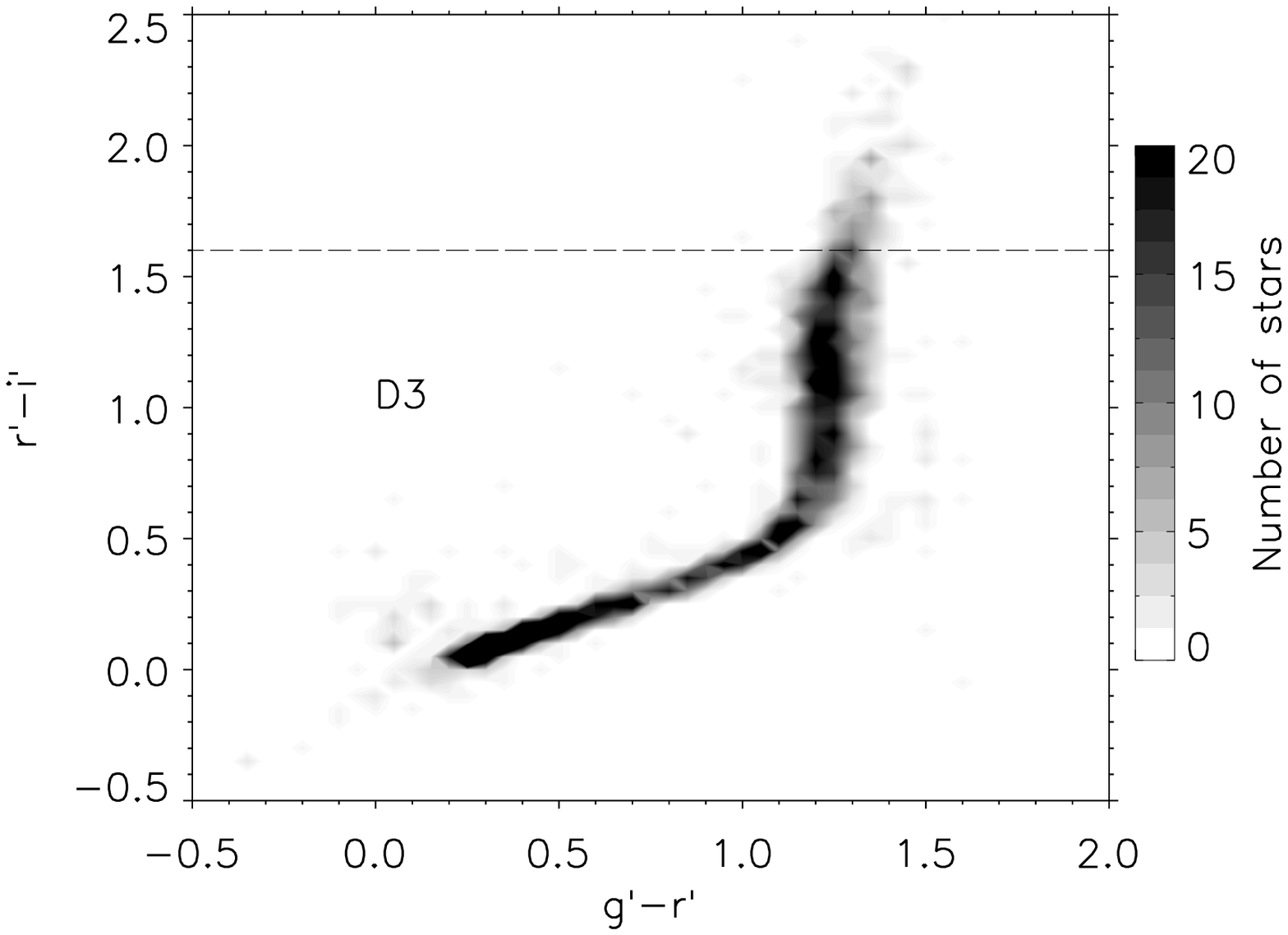} \epsfxsize=4.4cm \epsfbox[35 370 530 730]{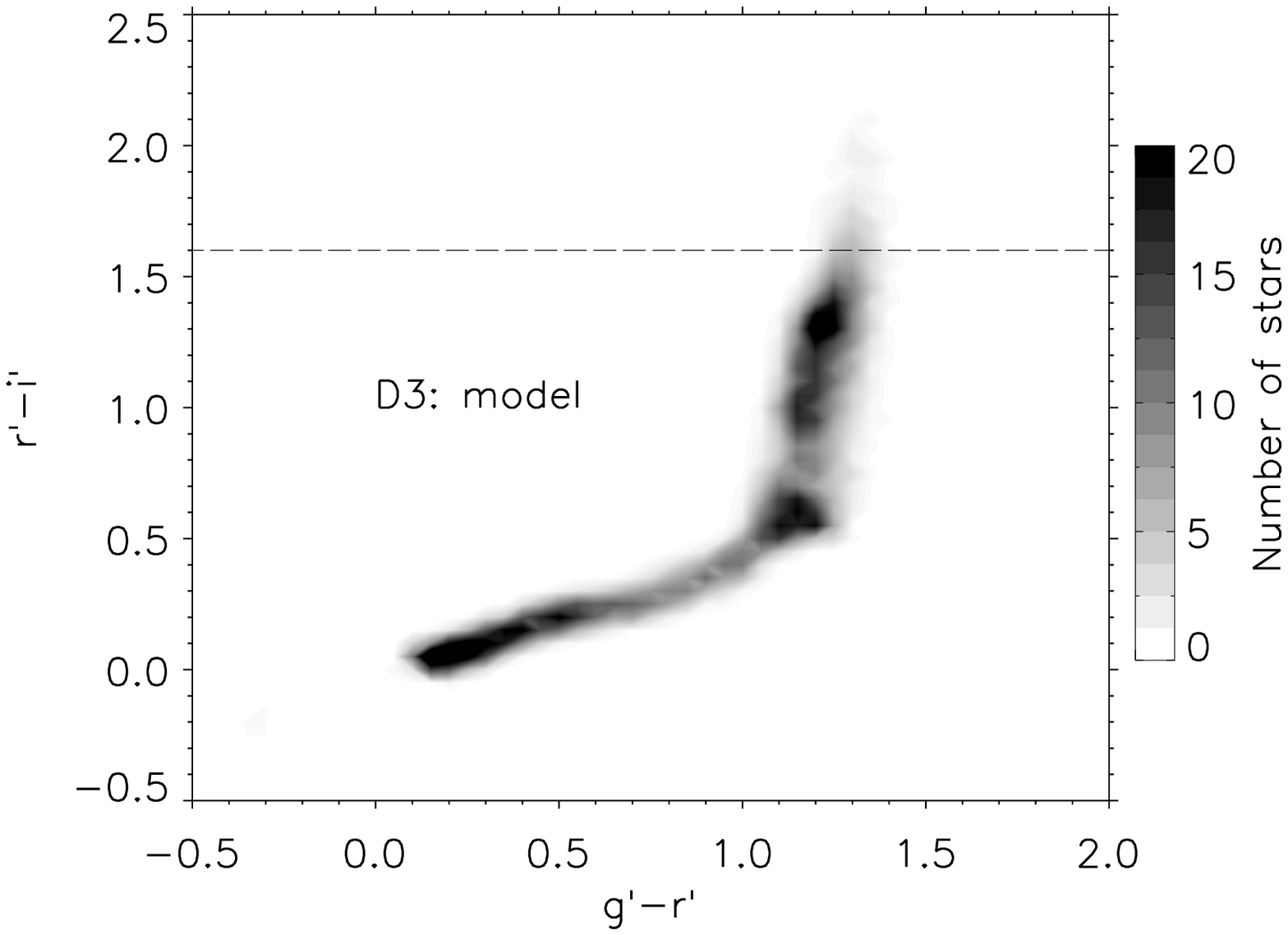}}

\caption{$g'-r'$ vs $r'-i'$ diagram of the D1, D2 and D3 field compared to  the synthetic colours predicted by the Besan\c{c}on model, all 
populations included. The greyscale indicates the number of stars per bin width of 0.05\,mag in each colour.}
\label{histall}

\end{figure}

As mentioned above, the stellar samples are contaminated by non-stellar
sources. As a large fraction of those fall inside the stellar locus and
stellar binaries and white dwarfs are also expected outside the main
sequence (see Fig.~2), only proper motions can be used to clean our
sample and remove galaxies and quasars. In the following all objects
classified as stellar are kept.

In Fig.~\ref{histD1} histograms in $g'-r'$, $r'-i'$
and $i'-z'$ for the D1 field are shown, with model predictions for each
population, thin disc, thick disc, and spheroid. Model predictions are
acceptable for all three populations, which are better separated in the
$r'-i'$ colour.

\subsection{Stellar densities}

Figure ~\ref{histall} shows the $g'-r'$ vs $r'-i'$ diagram for each
CFHTLS field, compared with model predictions, where the greyscale
indicates the number of stars.  The colour-colour diagrams in the three
fields are similar, and model predictions are in good general agreement
with the data. We notice, however, a few significant differences:

\begin{itemize}
\item In all three fields the predicted source density is a slightly
  low compared with the observations. The difference between the data
  and the predicted colour counts is of order $\sim$ 30\% for $r'-i'<
  0.3$.  We estimate that part of the excess in the data is
  overestimated, due to the contamination by galaxies, which in the
  blue is of order $\sim$ 13\% for the D1 field but larger for the D2
  and D3 field (see also Section 4).  On the other hand, the predicted
  counts of spheroid stars depend on three density parameters: the axis
  ratio, the power law exponent and the local normalization. These
  parameters have been already constrained by other data sets (Robin et
  al.  \cite{Robin2000}) but could be better adjusted using new large
  and deep surveys such as the CFHTLS or SDSS.  In order to perform
  better adjustments of these parameters, we need more fields with
  wider areas of the sky than those presented here.  The effects on the
  star counts of the derived IMF of this population as well as their
  density distribution can only be disentangled from large surveys in
  different areas of the sky. This will be considered in a future paper
  using the T0002 and T0003 releases.
  
\item In the range 0.3 $<$ $r'-i'$ $<$ 0.7 the thick disc population
  dominates while for $r'-i' > 0.8$ the thin disc dominates the counts.
  The whole colour range is well modeled in the three fields; the thick
  disc model fitted in Reyl\'e \& Robin (2001) fits well here too,
  in addition to the thin disc for 0.8 $<$ $r'-i'$ $<$ 1.4.

\item In the very red part of the diagram, at $r'-i'> 1.6$, the number
  of observed stars is significantly larger than predicted by the
  model. We do not expect large numbers of very high redshift galaxies
  and quasars at these colours. These point source objects are most
  probably disc M dwarfs. This implies that the assumed luminosity
  function of these stars has been underestimated in the model. We
  investigate this point further below.

\end{itemize}

\begin{figure}
\epsfxsize=9.0cm

\centerline{\epsfbox[60 350 520 640]{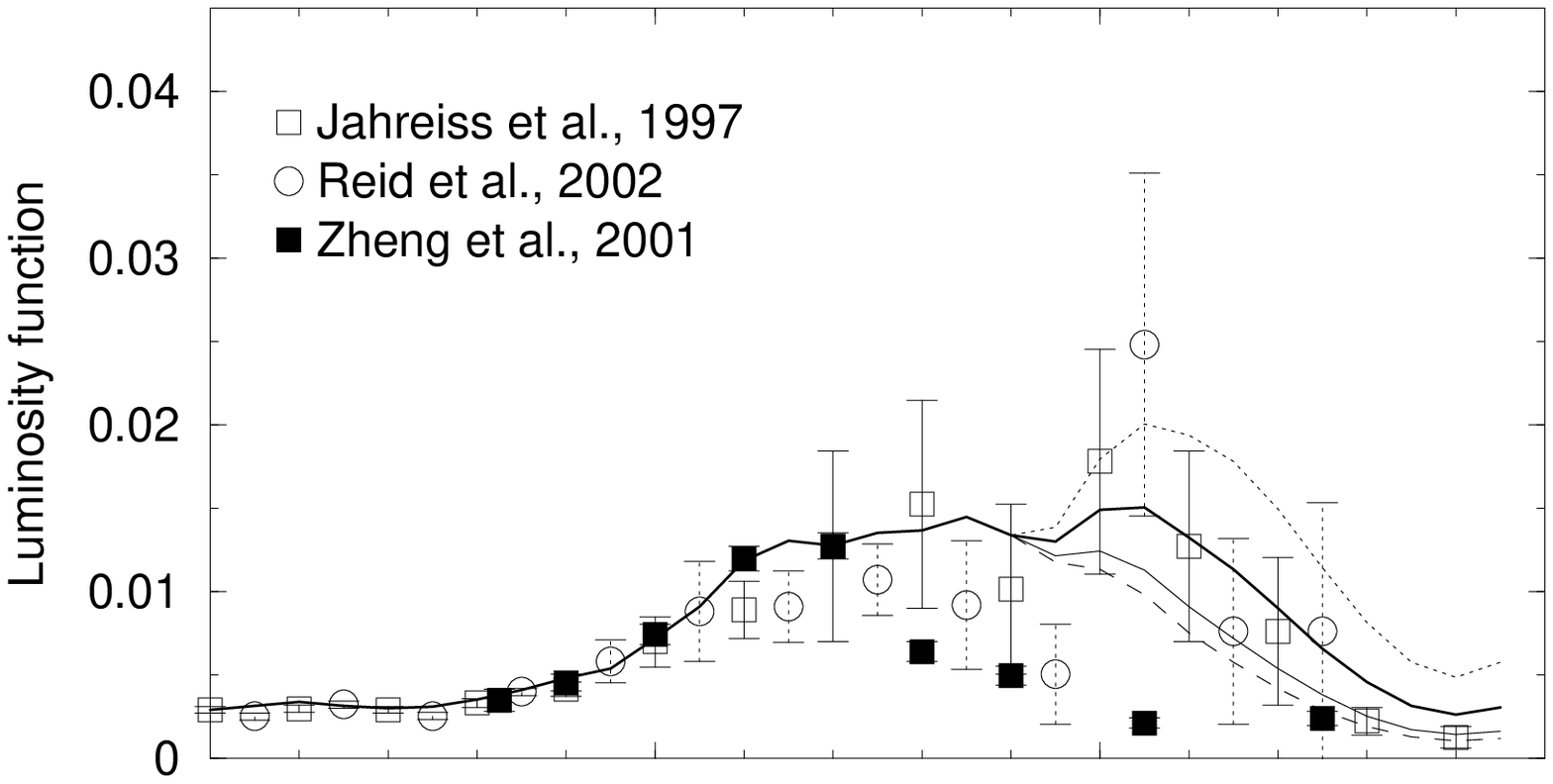}}
\vspace{-5cm} \hspace{1cm} $m_c$ = 0.15\Msun

\vspace{3.2cm}
 \centerline{\epsfbox[60 350 520 640]{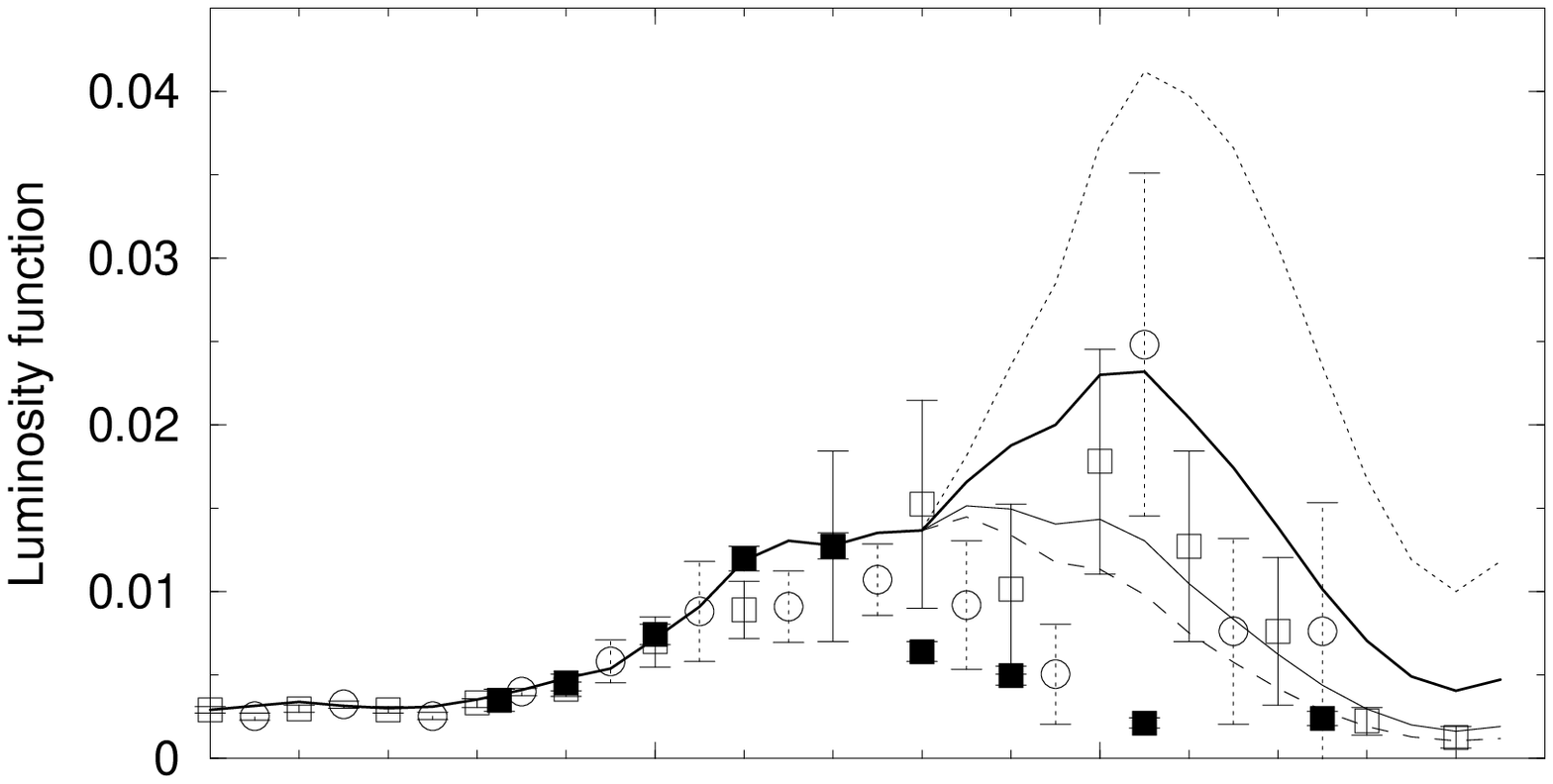}}
\vspace{-5cm} \hspace{1cm} $m_c$ = 0.20\Msun

\vspace{3.2cm}
 \centerline{\epsfbox[60 350 520 640]{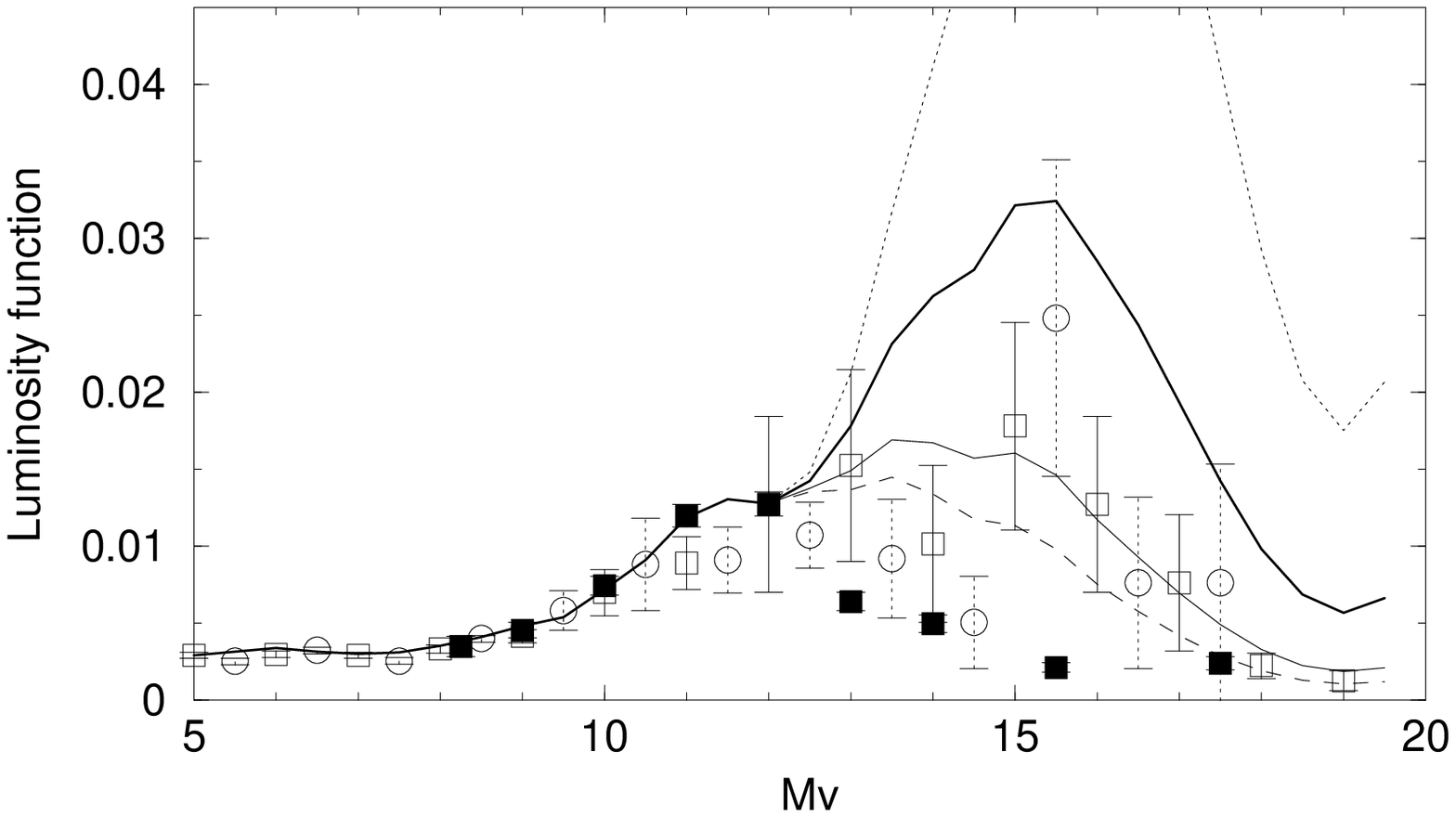}}
\vspace{-5cm} \hspace{1cm} $m_c$ = 0.25\Msun

\vspace{4.5cm}

\caption{Luminosity function in the V band in number of stars per $\rm pc^{3}$ per magnitude for different break points in mass. The top panel indicates  $m_{c} = 0.15$ \Msun, the middle panel $m_{c} = 0.20$ \Msun\ and the lower panel $m_{c} = 0.25$ \Msun.
  Open squares are from Jahreiss et al.'s (\cite{Jahreiss97})
  determination from the revised Catalogue of Nearby Stars, open
  circles are from Reid et al. (\cite{Reid2002}) determination using
  the PMSU survey combined with Hipparcos data, filled squares are from
  Zheng et al.'s (\cite{Zheng}) determination from the HST.  The lines
  show model luminosity functions assuming different slopes for the
  IMF: $\alpha=1.5$ is the standard Galaxy model (dashed line),
  $\alpha=2$ (thin line), $\alpha=3$ (thick line), and $\alpha=2$
  (dotted line) (see text).}
\label{lf}
\end{figure}

\subsection{The IMF at low masses}

The behaviour of the stellar luminosity function (LF) and mass
function (MF) for low-mass stars ($\rm < 1\,M_{\odot}$) is still under
debate.  Jahrei\ss\ et al. (\cite{Jahreiss97}) derived a local stellar
LF (within 20~pc) from the Catalogue of Nearby Stars revised with
Hipparcos data. Reid et al.  (\cite{Reid2002}) derived the nearby star
LF using the Palomar/Michigan State University sample (PMSU) combined
with the Hipparcos 25~pc sample. Both LFs are shown in Figure~\ref{lf}.
Error bars are large due to the small survey volumes.  For masses
smaller than 0.5\,$\rm M_{\odot}$, the determination of the MF is
hampered by the incompleteness of the different samples (Henry et al.
\cite{Henry97}, Chabrier \& Baraffe \cite{Chabrier2000}) and by the
unknown proportion of M dwarfs in binaries.  Chabrier
(\cite{Chabrier2003}) estimated that the mass function below $\rm
1\,M_{\odot}$ is consistent with a fraction of $\sim$ 50\% of M dwarf
binaries where 30 \% should have an M dwarf companion and 20\% a brown
dwarf secondary.

Also on Figure~\ref{lf} are superimposed the luminosity functions used
for simulations with our standard Galaxy model, as well as a few other
luminosity functions obtained by varying the IMF slope at low mass.
The luminosity function is made from segments of a power law IMF, as
given in equation~\ref{imf}, and a mass luminosity relation from
Delfosse et al.  (\cite{Delfosse2000}) in the magnitude range
12$<M_V<17$ and from theoretical models of Baraffe et al.
(\cite{Baraffe98}) at lower masses. The cutoff between absolute
magnitude 16 to 18 is mostly due to the mass-luminosity relation and
only weakly dependent on the IMF slope at the very low mass end.
However in the range 13$<M_V<16$ the luminosity function strongly
depends on the assumed IMF slope and on the mass at which the slope
changes:

\begin{equation}
\frac{dn}{dm} \propto m^{-\alpha}
\label{imf}
\end{equation}

with:\\
$\alpha$ = 1.5 for $m <$ 0.5 \Msun\ (standard Galaxy model)\\
$\alpha$ = 1.5 for $m <$ 0.5 \Msun\ and $\alpha$ = 2 for $m < m_{c}$\\
$\alpha$ = 1.5 for $m <$ 0.5 \Msun\ and $\alpha$ = 3 for $m < m_{c}$\\
$\alpha$ = 1.5 for $m <$ 0.5 \Msun\ and $\alpha$ = 4 for $m < m_{c}$\\
\medskip
and $ m_{c}= 0.15, 0.20, 0.25$ \Msun. 

\medskip

In the following the various tested IMFs are denoted LF$\alpha-m_{c}$, where alpha is the
IMF slope , and $m_{c}$ is the mass where the slope changes.

\begin{center}
\begin{table*}
\begin{tabular}{|c|c|ccccc|ccccc|ccccc|c|}
\hline
$r'-i'$&std&.15&.15&.15&.15&.15&.20&.20&.20&.20&.20&.25&.25&.25&.25&.25&data\\
     &              & 2 & 2.5 & 3 & 3.5 & 4 & 2 & 2.5 & 3 & 3.5 & 4 & 2 & 2.5 & 3 & 3.5 & 4 &\\
\hline
1.6 &     79 &   82 &   84 &   87 &   88 &   88 &   88 &  106 &  124 &  148 &  176 &  100 &  134 &  173 &  229 &   299 &        126\\
        $\sigma$&     4.2 &     3.9 &     3.7 &     3.5 &     3.4 &     3.4 &      3.4 &     1.8 &     0.2 &    --2.0 &    --4.5 &     2.3 &    --0.7 &    --4.2 &    --9.2 &   --15.4 & \\
\hline
1.8 &     25 &   30 &   35 &   38 &   41 &   50 &   35 &   47 &   62 &   76 &  102 &   41 &   58 &   84 &  121 &   172 &        57\\
        $\sigma$&     4.2 &     3.6 &     2.9 &     2.5 &     2.1 &     0.9 &      2.9 &     1.3 &    --0.7 &    --2.5 &    --6.0 &     2.1 &    --0.1 &    --3.6 &    --8.5 &   --15.2 & \\
\hline
2.0 &     11 &   13 &   16 &   18 &   23 &   26 &   15 &   18 &   29 &   42 &   59 &   15 &   24 &   37 &   61 &    95 &        21\\
        $\sigma$&     2.2 &     1.7 &     1.1 &     0.7 &    --0.4 &    --1.1 &      1.3 &     0.7 &    --1.7 &    --4.6 &    --8.3 &     1.3 &    --0.7 &    --3.5 &    --8.7 &   --16.1 & \\
\hline
\hline
tot&     115 &  125 &  135 &  142 &  151 &  165 &  138 &  171 &  215 &  266 &  337 &  157 &  216 &  294 &  411 &   566 &        204\\
        $\sigma$&     6.2 &     5.5 &     4.8 &     4.3 &     3.7 &     2.7 &      4.6 &     2.3 &    --0.8 &    --4.3 &    --9.3 &     3.3 &    --0.8 &    --6.3 &   --14.5 &   --25.3 & \\
\hline
\end{tabular}
\label{table_lf_sigma_d1}
\caption{Number of stars in the D1 field for $\rm i' < 21$ and in different
intervals of  $r'-i'$. Column 1 indicates the middle of the interval of width 0.2, 
``tot'' means the total of the 3 interval considered, that is $1.5< r'-i'<2.1$. Columns 
2 to 17 give the model counts for each of
the tested LF while column 18 contains the observed counts. Models are described by two 
parameters: on the first line of the column heading the value of the mass of changing 
slope is indicated, the second line gives the value of $\alpha$ (see eq.\ref{imf}).
The column values include, in the first line,  the number of stars in the colour interval, 
the second line gives the difference between the data and simulated counts in number of 
sigmas, assuming that the noise is
dominated by the Poisson statistics (see text). Model with values between -3 and +3 are 
considered as acceptable in the colour bin.}

\end{table*}
\end{center}

\begin{center}
\begin{table*}
\begin{tabular}{|c|c|ccccc|ccccc|ccccc|c|}
\hline
$r'-i'$&std&.15&.15&.15&.15&.15&.20&.20&.20&.20&.20&.25&.25&.25&.25&.25&data\\
     &              & 2 & 2.5 & 3 & 3.5 & 4 & 2 & 2.5 & 3 & 3.5 & 4 & 2 & 2.5 & 3 & 3.5 & 4 &\\
\hline
1.6 &         80 &   75 &   81 &   81 &   87 &   89 &   91 &  106 &  122 &146 &  167 &  101 &  134 &  176 &  235 &   310 &        141\\
        $\sigma$&     5.1 &     5.6 &     5.1 &     5.1 &     4.5 &     4.4 &      4.2 &     2.9 &     1.6 &    --0.4 &    --2.2 &     3.4 &     0.6 &    --2.9 &    --7.9 &   --14.2 &\\
\hline
1.8 &         28 &   29 &   32 &   35 &   39 &   48 &   33 &   40 &   56 & 70 &   94 &   36 &   53 &   76 &  112 &   156 &        49\\
        $\sigma$&     3.0 &     2.9 &     2.4 &     2.0 &     1.4 &     0.1 &      2.3 &     1.3 &    --1.0 &    --3.0 &    --6.4 &     1.9 &    --0.6 &    --3.9 &    --9.0 &   --15.3 &\\
\hline
2.0 &          9 &   10 &   14 &   17 &   20 &   24 &   14 &   18 &   26 & 36 &   49 &   13 &   23 &   33 &   55 &    85 &        21\\
        $\sigma$&     2.6 &     2.4 &     1.5 &     0.9 &     0.2 &    --0.7 &      1.5 &     0.7 &    --1.1 &    --3.3 &    --6.1 &     1.7 &    --0.4 &    --2.6 &    --7.4 &   --14.0 &\\
\hline
\hline
tot &  116 &  115 &  126 &  132 &  146 &  160 &  138 &  164 &  203 &  253 &  309 &  150 &  211 &  285 &  402 &   552 &        221\\
        $\sigma$&     7.1 &     7.1 &     6.4 &     6.0 &     5.0 &     4.1 &      5.6 &     3.8 &     1.2 &    --2.2 &    --5.9 &     4.8 &     0.7 &    --4.3 &   --12.2 &   --22.3 &\\
\hline
\end{tabular}
\label{table_lf_sigma_d2}
\caption{Same as Table \ref{table_lf_sigma_d1} but for field D2.}

\end{table*}
\end{center}
\begin{center}
\begin{table*}
\begin{tabular}{|c|c|ccccc|ccccc|ccccc|c|}
\hline
$r'-i'$&std&.15&.15&.15&.15&.15&.20&.20&.20&.20&.20&.25&.25&.25&.25&.25&data\\
     &              & 2 & 2.5 & 3 & 3.5 & 4 & 2 & 2.5 & 3 & 3.5 & 4 & 2 & 2.5 & 3 & 3.5 & 4 &\\
\hline
1.6 &     75 &   71 &   76 &   76 &   78 &   81 &   81 &  100 &  116 &  141 &  167 &   93 &  125 &  160 &  208 &   288 &        126\\
        $\sigma$&     4.5 &     4.9 &     4.5 &     4.5 &     4.3 &     4.0 &      4.0 &     2.3 &     0.9 &    --1.3 &    --3.7 &     2.9 &     0.1 &    --3.0 &    --7.3 &   --14.4 & \\
\hline
1.8 &     22 &   25 &   31 &   34 &   39 &   44 &   31 &   40 &   54 &   69 &   93 &   36 &   51 &   75 &  111 &   161 &        57\\
        $\sigma$&     4.6 &     4.2 &     3.4 &     3.0 &     2.4 &     1.7 &      3.4 &     2.3 &     0.4 &    --1.6 &    --4.8 &     2.8 &     0.8 &    --2.4 &    --7.2 &   --13.8 & \\
\hline
2.0 &     10 &   12 &   15 &   18 &   22 &   26 &   14 &   18 &   26 &   36 &   54 &   15 &   24 &   39 &   60 &    96 &        30\\
        $\sigma$&     3.7 &     3.3 &     2.7 &     2.2 &     1.5 &     0.7 &      2.9 &     2.2 &     0.7 &    --1.1 &    --4.4 &     2.7 &     1.1 &    --1.6 &    --5.5 &   --12.0 & \\
\hline
\hline
tot &    107 &  108 &  122 &  128 &  138 &  151 &  125 &  158 &  196 &  246 &  315 &  142 &  200 &  274 &  379 &   545 &        219\\
        $\sigma$&     7.6 &     7.5 &     6.6 &     6.1 &     5.5 &     4.6 &      6.4 &     4.1 &     1.6 &    --1.8 &    --6.5 &     5.2 &     1.3 &    --3.7 &   --10.8 &   --22.0 & \\
\hline
\end{tabular}
\label{table_lf_sigma_d3}
\caption{Same as table \ref{table_lf_sigma_d1} but for field D3.}

\end{table*}
\end{center}

We note that simulated stars considered here are single stars. Distant
binary systems may be not correctly identified as stars, although from
a detailed analysis of the binary effect (a complete analysis of the
binary effect is postponed to the next paper of this series), we have
estimated that at $r'-i'>1.6$ (that is M$_V>14$) the correction for
stars missed for this reason is negligible. At $r'-i'<1.5$ the
correction would be less than 25\% and at $1.5 < r'-i' <1.6$ it is less
than 13\%.  Hence with respect to the very faint end of the LF studied
here, the binary effect is expected to be negligible because these
stars are too close to have been missed even in binary systems. We
limit our further comparison to $r'-i'>1.5$.

\begin{figure}
\epsfysize=11.5cm
\centerline{\epsfbox[40 20 570 750]{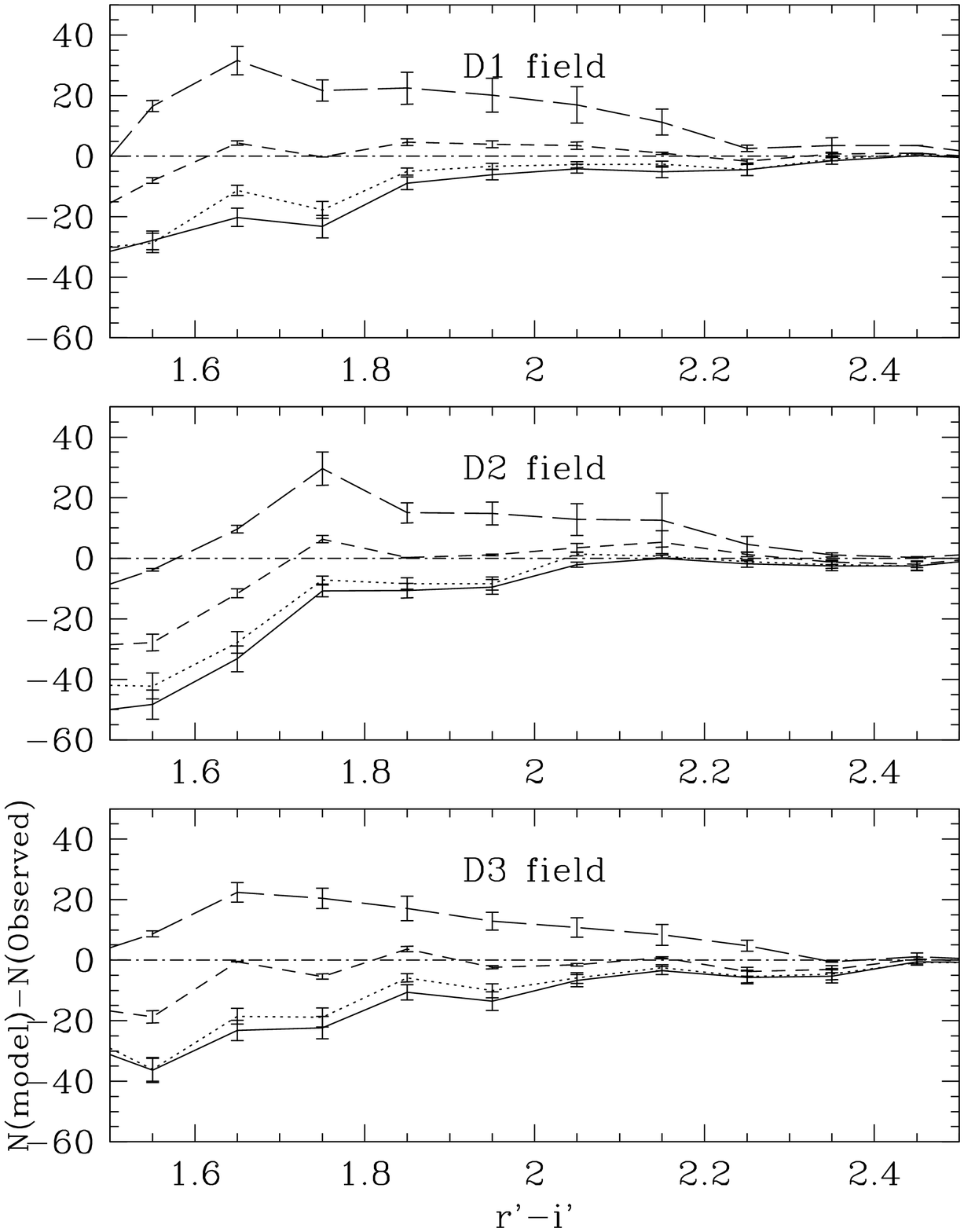}}
\caption{Difference between model and data as a function  of the $r'-i$ colour for the D1, D2 and D3 field.
 The solid line indicates the standard model ($\alpha=1.5$) with $\rm m_{c} = 0.20$ \Msun, the dotted line  assuming $\alpha=2$, the short dashed and the long dashed line assuming $\alpha=3$  and $\alpha=4$  respectively, as explained in the text. The error bars are calculated assuming that the noise is due to the Poisson statistics.   }
\label{hist_newlf}

\end{figure}

Figure~\ref{histall} shows clearly a deficit of late type thin disc
dwarfs in model predictions at $r'-i'> 1.6$, for the standard Galaxy
model.  We have attempted to fit the LF to the available CFHTLS data in
the 3 fields using the IMF formula from eq. (1).
Figure~\ref{hist_newlf} shows the difference of the predicted star
counts compared to observations for the three tested IMFs: $\alpha=2$ ,
$\alpha=3$ and $\alpha=4$ with $\rm m_{c} = 0.20$ \Msun.

In Table 4, 5 and 6 we give the number of stars in $r'-i'$ intervals
from the D1, D2 and D3 field respectively, and, for comparison, the
numbers simulated from tested LFs, varying $m_{c}$ from 0.15 to 0.25
and $\alpha$ at $m<m_{c}$ from 1.5 to 4. The standard deviation of the
models relative to the data are estimated from Poisson statistics:
assuming that the main source of error in the data is Poisson noise, we
estimate the deviation of the model by computing :

\[\sigma_{model} = \frac{N_{data}-N_{model}}{\sqrt{N_{data}}}\]

which gives the relative difference in the counts in units of the
Poissonian scatter. Acceptable models have $\sigma_{model}$ in the range
[$-3,3$].

 From these Tables we can conclude that :
\begin{itemize}

\item The standard LF (with $\alpha$ = 1.5, i.e. no change of slope) is
  rejected at more than 6 sigmas in the three fields.
  
\item The LF with a $m_{c}= 0.15$ \Msun\ is deficient in stars compared
  with the data in all three fields, regardless of the slope.
  
\item Consistency with the data can be obtained with $m_{c}=0.25$ or
  $m_{c}=0.20$ \Msun. Depending on which $m_{c}$ we take, the value of
  $\alpha$ is slightly changed. Having a higher $m_{c}$ allows a
  smaller $\alpha$. Constraints from the local LF (see Figure 11) also
  indicate that having $m_{c} > 0.25$ \Msun\ would not allow large
  values for $\alpha$.

\item The counts in all 3 fields are best fitted by an IMF slope of
  $\alpha=3$ (with $ m_{c}=0.20$ \Msun) or $\alpha=2.5$ (with
  $m_{c}=0.25$ \Msun).  As seen from Figure 11 this LF is in acceptable
  agreement with the local LF as determined from the 25 pc sample.
  Comparing this IMF (either with $m_{c} = 0.20$ or 0.25 \Msun\.) with
  Chabrier (\cite{Chabrier2003}) log-normal LF, yields good agreement
  but the revised IMF from Chabrier (\cite{Chabrier2004}) gives a too
  few low mass stars for 0.10-0.15 \Msun\ compared with our
  determination and with the local IMF.
  We emphasize that we have confirmed, from our intermediate distance
  sample that the disc IMF does not decrease for masses above the 0.1
  \Msun\ limit.

\end{itemize}
It should be noted that the number of stars decreases rapidly
with the increasing $r'-i'$ colour. A slight change in the zero point
of the photometric calibration in $r'$ or $i'$ may change slightly our
conclusions. We estimate from Tables 4, 5 and 6 that a systematic shift
of 0.05 magnitude in $r'-i'$ would produce a change of about 0.5 on the
slope $\alpha$. However such a shift is improbable as it would be seen
also at the blue end of the sequence ($r'-i'=0$) which is not the case,
as seen in Figure 9.

\subsection{Comparison with HST results}
Zheng et al. (\cite{Zheng}) determined the luminosity function from a
sample of about 1400 M dwarfs in 148 fields using the WFC2 and 162
fields from PC1 with the HST.  Their sample is characterized by a mean
height above the plane of 1.5 kpc, with very few stars at vertical
height $<$1 kpc. They derive their LF and IMF taking into account a
probable metallicity gradient, by adopting a metallicity of --0.5 dex
at 1.5 kpc, and a colour-absolute magnitude relation varying with
metallicity. Hence their sample is dominated by what we usually call
the thick disc population. They deduce an IMF slope of $\alpha=-0.10$
or $\alpha=-0.47$ with or without the metallicity gradient taken into
account.

The sample considered in this paper is significantly different from the
HST sample, as it is dominated by stars at distances above the plane of
150 to 450 pc with a mean distance of 350 pc for stars at $r'-i'=1.6$
and 210 pc for stars having $r'-i'=2.0$. This has two consequences: 1)
the sample is less biased by unresolved binaries and 2) it is dominated
by the normal thin disc population and more comparable with the local
sample which is used to determine the LF in the solar neighbourhood
(Reid et al. \cite{Reid2004}).

Reyl\'e \& Robin (\cite{Reyle2001}) have performed the first
determination of the thick disc IMF from a multi-directional analysis
of star counts. They obtained an IMF $dN/dm \propto m^{-0.5}$ in the
mass range $0.2<m<0.8\Msun$, which is in agreement with the IMF deduced
by Zheng et al. (\cite{Zheng}) from the HST sample, reinforcing the
ideas that: firstly Zheng et al. (\cite{Zheng}) have measured the thick
disc IMF, rather than the thin disc one; secondly the thin disc and
thick disc have different IMF slopes at low masses. The IMF found by
Reyl\'e \& Robin (\cite{Reyle2001}) in the thick disc is well in
agreement with the one determined in globular clusters (Paresece \& De
Marchi \cite{Paresece}) and significantly different from the one found
in the local thin disc (Kroupa \cite{Kroupa2001}).

The origin of the thick disc has long been a matter of debate. Nowadays
favoured scenarii explain the thick disc by one or more accretions of
galaxy satellites at early epochs of the Galaxy's formation, or by star
formation from gas accreted during a chaotic period of hierarchical
clustering (Brook et al. \cite{Brook2004}). The thick disc is old and
metal poor relative to the sun and it is also enhanced in alpha
elements. Abundance determinations (Gratton et al. \cite{Gratton2000})
also show that there has been a discontinuity in the star formation
between the thick disc and the thin disc of at least 1 Gyr. The
conditions of star formation at the epoch of thick disc formation were
clearly different from the present conditions in the thin disc.  Larson
(2005) has analysed the physical conditions required for the thermal
coupling of gas and dust in cloud fragmentation. He studied the roles
of the metallicity, background radiation and dust environment on the
Jeans mass, hence on the typical mass of the stars formed. The combined
effects of the metallicity and the possible lack of dust at
cosmological epochs could increase the peak mass of the IMF relative to
the present one. Moreover as the cosmic background temperature was
higher in the past, a higher minimum cloud temperature exists, which
also implies a higher Jeans mass. These conditions may well explain the
fact that the IMF found in the thick disc has a typical mass higher
than the thin disc, and is deficient in the very low mass stars which
are found in the present disk MF.

\section{Conclusions}

We have presented an analysis of the stellar populations in the CFHTLS
using catalogues and images from the first public data release. Our
population synthesis approach allowed us to test stellar libraries and
to identify different stellar types and Galaxy components using
colour-colour diagrams. We discuss the locations of various stellar
species such as white dwarfs, late-type and brown dwarfs and binary
systems in the MEGACAM filter/detector combination.  The contamination
of the stellar sample by quasars and compact galaxies is quantified
using spectroscopic data from the VIMOS-VLT Deep Survey (VVDS). The
percentage of the galaxy contamination depends very much on the $r'-i'$
colour and can reach a maximum of 23.5\% for $i' < 22.0$ and $r'-i' <
0.5$.
  
Our main conclusions concern the luminosity and mass function (MF) at
low mass for the disc population.  This data set favours an MF slope of
$\alpha=2.5 \pm 1.0$ for $m< 0.25$ \Msun\ or $\alpha=3.0 \pm 1.0$ for
$m<$0.2 \Msun\, which although steep compared with previous
investigations from other deep imaging surveys (such as Zheng et al.
who used HST images) is still in agreement with local determinations of
the IMF. This discrepancy can be explained by differences in the mean
age and physical conditions of star formation of the samples, one being
at about 1 kpc or more where the thick disc population is expected to
dominate, and ours being at 150-450 pc and dominated by the thin disc.
This discrepancy between the thin disc and thick disc IMFs could be
explained if for physical reasons (for example, lack of dust, higher
temperature backgound radiation or metallicity) very low mass star
formation has been less efficient at the epoch of the thick disc
formation.

The new IMF as determined here cannot be extrapolated to masses below
0.1 \Msun.  It is probable from the numbers of known brown dwarfs in
clusters that the IMF starts to decrease near the H burning limit
(Kroupa \cite{Kroupa2001}).

In future papers we plan a more detailed analysis of these stellar
populations, in particular the IMF at low masses of the disk, thick
disk and spheroid and the old population density distribution up to
several tens of kiloparsecs. 

This might be performed using more accurate star-galaxy separation and
by accounting for binary frequency in the modelling.  The combination
of multiband wide survey coverage together with proper motions will
enable us to count thick disc and halo white dwarfs, and to constrain
on the fraction of baryonic dark matter present in the form of stellar
remnants.

\acknowledgements{We would like to warmly thank Roser Pello for
  providing us with simulated galaxy colours in the MEGACAM photometric
  systems for a range of different redshifts, Pierre Bergeron for
  providing his white dwarf models and Thibault Lejeune for his
  invaluable assistance in the computation of the synthetic colours
  from the stellar libraries.  We would like also to thank the VVDS
  consortium for providing us with spectroscopic observations.  Many
  thanks also to Bernard Debray who is responsible for providing the
  web interface for the Besancon Galaxy model.  MS is supported by an
  APART fellowship.}

{}

\end{document}